\documentclass[fleqn,usenatbib]{mnras}

\usepackage{newtxtext,newtxmath}

\usepackage[T1]{fontenc}

\DeclareRobustCommand{\VAN}[3]{#2}
\let\VANthebibliography\thebibliography
\def\thebibliography{\DeclareRobustCommand{\VAN}[3]{##3}\VANthebibliography}

\usepackage{graphicx}	
\usepackage{amsmath}	
\usepackage{multirow}

\title[Carbon chains in L1544 and IRAS 16293]{Carbon chain diversity in L1544 and IRAS 16293-2422: an astrochemical pathfinder study for the SKAO} 

\author[Giani et al.]{
Lisa Giani,$^{1}$\thanks{E-mail: lisa.giani@inaf.it}
Eleonora Bianchi,$^{1}$
Anthony Remijan,$^{2}$ 
Claudio Codella,$^{1,3}$
Giovanni Sabatini,$^{1}$
\and
Linda Podio,$^{1}$
Cecilia Ceccarelli,$^{3}$
Marta De Simone,$^{4}$
Nadia Balucani,$^{5}$
Paola Caselli,$^{6}$
\and
Eric Herbst,$^{7,8}$
Francois Lique,$^{9}$ 
Silvia Spezzano,$^{6}$
Charlotte Vastel$^{10}$
and Brett McGuire$^{11,12}$
\\\\
$^{1}$INAF, Osservatorio Astrofisico di Arcetri, Largo E. Fermi 5, I-50125, Firenze, Italy\\
$^{2}$National Radio Astronomy Observatory, Charlottesville, VA 22903, USA\\
$^{3}$Univ. Grenoble Alpes, CNRS, IPAG, 38000 Grenoble, France\\
$^{4}$ European Southern Observatory, Karl-Schwarzschild Str. 2, 85748 Garching bei M\"unchen, Germany\\
$^{5}$Dipartimento di Chimica, Biologia e Biotecnologie, Via Elce di Sotto 8, 06123 Perugia, Italy \\
$^{6}$Max-Planck-Institut f\"ur extraterrestrische Physik (MPE), 
Giessenbachstrasse 1, 85748 Garching, Germany\\
$^{7}$Department of Chemistry, University of Virginia, Charlottesville, VA 22904, USA\\
$^{8}$Department of Astronomy, University of Virginia, Charlottesville, VA 22904, USA\\
$^{9}$Université du Rennes, CNRS, IPR (Institut de Physique de Rennes)UMR 6251, F-35000 Rennes, France\\
$^{10}$IRAP, Université de Toulouse, CNRS, UPS, CNES, 31400 Toulouse, France\\
$^{11}$Department of Chemistry, Massachusetts Institute of Technology, Cambridge, MA 02139, USA\\
$^{12}$National Radio Astronomy Observatory, Charlottesville, VA 22903, USA
}

\date{Accepted XXX. Received YYY; in original form ZZZ}

\pubyear{\the\year{}}

\begin{document}
\label{firstpage}
\pagerange{\pageref{firstpage}--\pageref{lastpage}}
\maketitle

\begin{abstract}

Astrochemical observations have revealed a surprisingly high level of chemical complexity, including long carbon chains, in the earliest stages of Sun-like star formation. The origin of these species and whether they undergo further growth, possibly contributing to the molecular complexity of planetary systems, remain open questions. 
We present recent observations performed using the 100-m Green Bank Telescope of the prestellar core L1544, and the protostellar system IRAS 16293–2422. In L1544, we detected several complex carbon-bearing species, including C$_2$S, C$_3$S, C$_3$N, c-C$_3$H, C$_4$H and C$_6$H, complementing previously reported emission of cyanopolyynes. In IRAS 16293–2422, we detected c-C$_3$H and, for the first time, HC$_7$N. 
Thanks to the high spectral resolution, we refine the rest frequencies of several c-C$_3$H and C$_6$H transitions.
We perform radiative transfer analysis, highlighting a chemical difference between the two sources: IRAS 16293–2422 shows column densities 10 to 100 times lower than L1544. 
We perform astrochemical modeling, employing an up-to-date chemical network with revised reaction rates.
Models reproduce the general trends, with cyanopolyyne and polyynyl radical abundances decreasing as molecular size increases, but underestimate the abundances of cyanopolyynes longer than HC$_5$N by up to two orders of magnitude. 
Current models, which include the dominant neutral–neutral formation routes, cannot account for this discrepancy, suggesting that the chemical network is incomplete.
We propose that additional ion–molecule reactions are crucial for the formation of these species. Developing a more comprehensive chemical network for long carbon chains is essential for accurately interpreting present and future observations.

\end{abstract}




\section{Introduction}

The formation of a Solar-type star and its planetary system begins with a prestellar core, a dense ($\geq$10$^{5}$ cm$^{-3}$) and cold ($\sim$10 K) condensation of gas typically spanning $\sim$ 10$^3$--10$^4$ au which eventually collapses under its own gravity \citep[e.g.][]{BensonMyers1989, Bergin2007,Keto2008}. As the collapse proceeds, a protostar forms in the central region, heating the gas and generating shocks in the surrounding medium \citep[e.g.][]{Stahler&Palla}. A surprisingly rich chemistry has been observed even in the initial stages of this process, marked by the gas-phase detection of complex molecules relevant for the formation of prebiotic species, such as interstellar complex organic molecules \citep[iCOMs, primarily C-bearing species with at least 6 atoms; e.g.,][and references therein]{Ceccarelli2004, Herbst2009,Ceccarelli2023}. The prestellar phase is particularly rich in deuterated species and iCOMs, which are primarily observed at (sub-)millimeter wavelengths \citep{Crapsi2005, Caselli2012, Vastel2014,Vastel2019,Bizzocchi2014,Jimenez2016,Spezzano2017,Punanova2018, Scibelli2024}. Furthermore, recent radio observations of the starless core TMC-1, conducted with the Green Bank Telescope (GBT\footnote{https://greenbankobservatory.org/science/telescopes/gbt/}) and the YEBES 40-m\footnote{https://rt40m.oan.es/}, have also revealed an active chemistry of complex carbon-bearing species, including cyclic hydrocarbons such as c-C$_9$H$_8$, benzonitrile (c-C$_6$H$_5$CN), cyanopyrene (C$_{16}$H$_{9}$CN) and cyanocoronene (C$_{24}$H$_{11}$CN) \citep[e.g.][and references therein]{McGuire2020,Cernicharo2021,Siebert2022,Sita2022,Agundez2023,Silva2023,Cernicharo2023,Fuentetaja2023,Remijan2023,2025wenzel,2025wenzel2}. This peculiar chemistry is still mostly unexplored due to limitations in current radio facilities, but there are hints that it may not be confined to TMC-1 and could be widespread in starless and prestellar cores \citep{Burkhardt2021b}. On the other hand, a dichotomy in chemical composition has been revealed in protostellar sources. This is characterized by two extreme cases: hot corinos and Warm Carbon Chain Chemistry (WCCC) sources. Hot corinos are regions approximately 100 au in size, where temperatures reach at least 100 K, causing dust mantles to sublimate and, in turn, increasing the abundance of iCOMs \citep{Ceccarelli2004, Ceccarelli2023}. Conversely, WCCC sources are poor in iCOMs but enriched in unsaturated small carbon chains \citep[such as c-C$_3$H$_2$,][]{Sakai2008,Sakai2013,Taniguchi2024} in regions that can extend up to 1000 au, where temperatures exceed 20 K and CH$_4$ sublimates.
The origin of this chemical diversity remains an open question, and the recent discovery of hybrid sources, such as B335, L483 and IRAS 15398-3359, further complicates the picture \citep{Imai2016, Oya2017, Jacobsen2019, 2023okoda}. While a different duration of the prestellar stage or external UV illumination have been proposed as potential explanations for the varying chemistry, the limited number of observations of complex carbon chains prevents a comprehensive understanding of this dichotomy \citep{Bouvier2020, Bouvier2022}. 
In this work, we present a spectral survey conducted with the GBT in the Ku (13.5 - 15.4~GHz) and X (8.0 - 11.6~GHz) bands, targeting one of the prototypical prestellar cores, L1544, along with the molecular envelope surrounding representative protostars embedded within hot corinos, specifically IRAS 16293-2422. The subsequent sections are organized as follows: Section \ref{sec:sample} details the two sources, Section \ref{sec:obs} describes the observations, Section \ref{sec:results} presents the results, and Section \ref{sec:discussion} discusses the origin of the observed chemical diversity. Finally, Section \ref{sec:conclusions} provides the conclusions.

\begin{figure}
\includegraphics[scale=0.6]{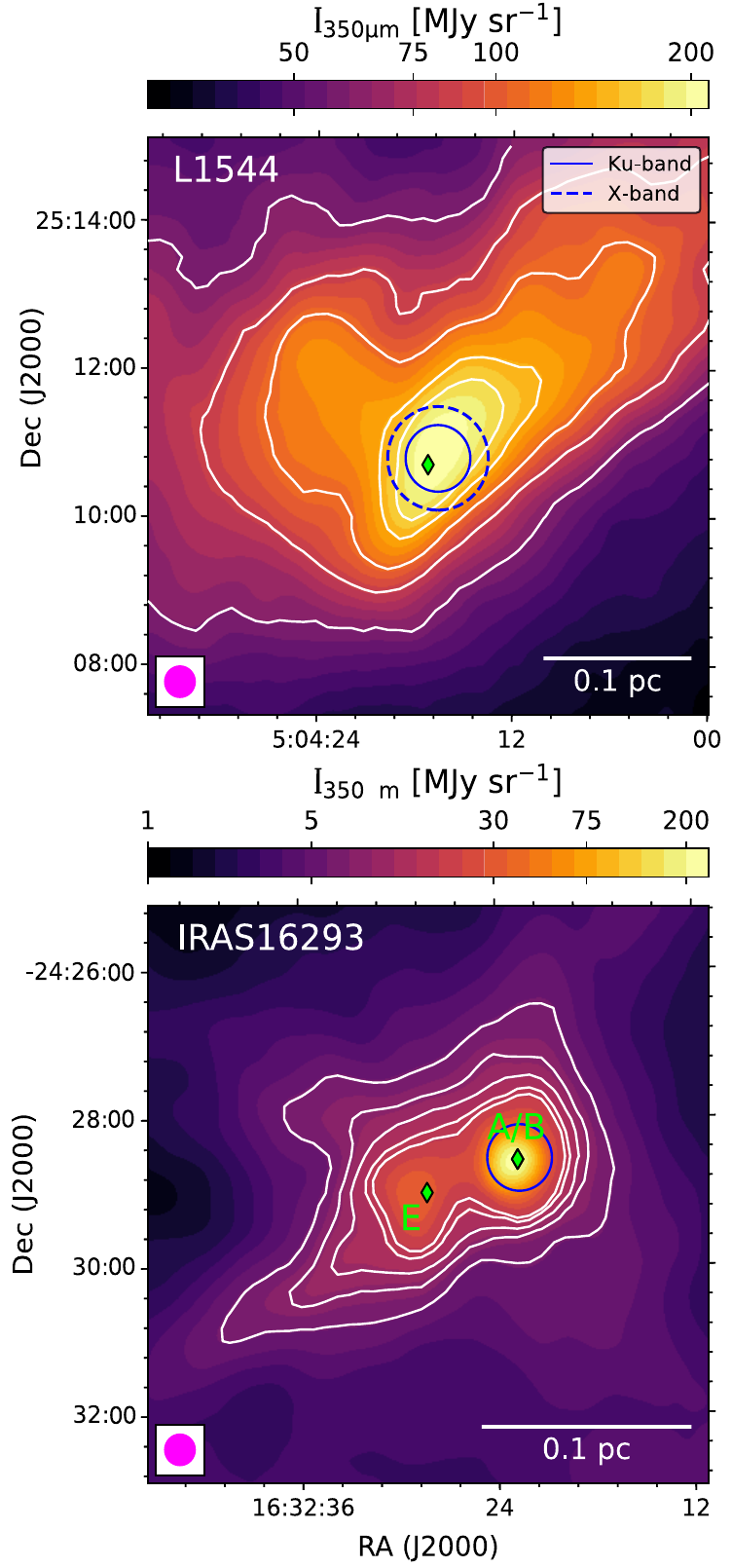}
  \caption{Overview of the L1544 (upper panel) and IRAS 16293--2422 (lower panel) regions as traced by the \textit{Herschel}\protect\footnotemark[3] continuum maps at 350~$\mu$m. The white contours correspond to [5,7,9,12,14]$\sigma$, where $\sigma$ is 12~MJy~sr$^{-1}$ in L1544, and 1.1~MJy~sr$^{-1}$ in IRAS 16293--2422, respectively. The FoVs of the GBT observations are shown as blue circles -- solid line for Ku-band and dashed line for X-band observations (see Sect.~\ref{sec:obs}). Green diamonds identify the continuum ($\sim$1.3~mm) peak position in L1544  \citep{WardThompson99} and the position of the A/B, and E objects in IRAS 16293--2422 \citep{Kahle2023}. The size of the \textit{Herschel} beam ($25\arcsec$) is shown in the bottom left corner, while the scale bar is shown in the bottom right.}
  \label{fig:Herschelmaps}
\end{figure}
\footnotetext[3]{https://archives.esac.esa.int/hsa/whsa/}

\begin{figure}
\includegraphics[scale=0.57]{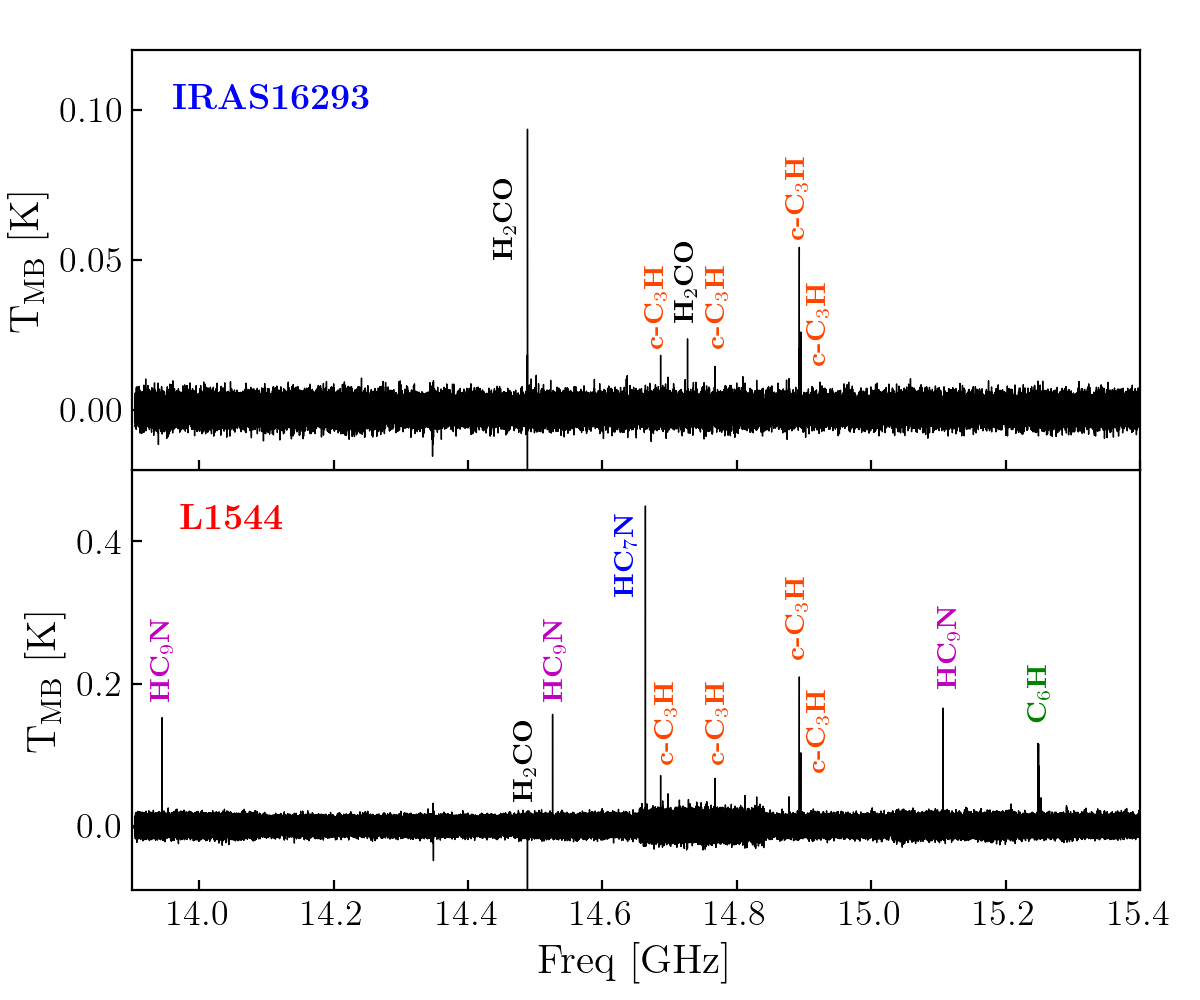}
  \caption{Full continuum-subtracted spectra (in main-beam temperature, T$_{\rm MB}$) of the IRAS 16293-2422 envelope (Upper panel) and the L1544 prestellar source (Lower panel) observed in Ku Band (13.9 -- 15.41 GHz). The brightest lines associated with the detected species (see Tab. \ref{Tab:lines} for the list of all the detections) are labelled. The L1544 spectrum has a spectral resolution of 1.4 kHz ($\sim$ 30 m s$^{-1}$), while the IRAS 16293 spectrum has been here smoothed to 12 kHz ($\sim$240 m s$^{-1}$) to increase the S/N of the emission lines. In the L1544 spectrum, the H$_2$CO line at $\sim$14.5 GHz is observed in absorption. }
  \label{fig:full-band-X}
\end{figure}

\begin{figure*}
\includegraphics[scale=0.61]{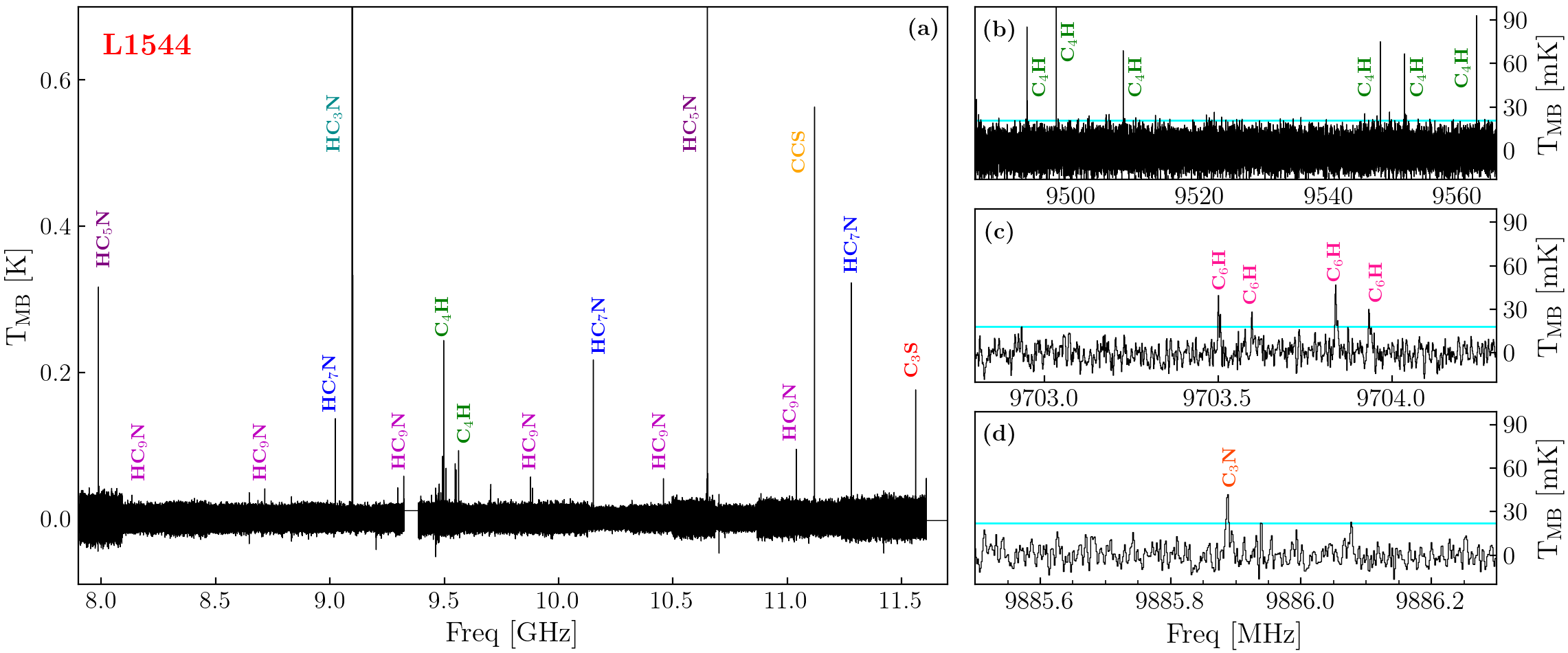}
  \caption{(Panel a): Full spectrum (in main-beam temperature, T$_{\rm MB}$) of the L1544 prestellar source observed in X Band (7.8 -- 11.6 GHz). The brightest lines associated with the detected species (see Tab. \ref{Tab:lines} for the list of all the detections) are labelled. The spectral resolution is 2.8 kHz ($\sim$ 60 m s$^{-1}$). (Panel b, c, d):. Zooms in to highlight selected lines of C$_4$H, C$_6$H and C$_3$N, respectively. The cyan solid lines correspond to 3$\sigma$.
  }
  \label{fig:full-bands-Ku}
\end{figure*}

\section{The sample}\label{sec:sample}
We chose as targets two well-studied sources that are prototypical of their evolutionary stage: L1544, a prestellar core, and IRAS 16293-2422, a multiple protostellar system. For both sources, their physical structure is well-known, and they are recognized for hosting a rich chemistry, making them ideal targets for exploring molecular composition at low frequencies.


\subsection{The L1544 prestellar core} 

L1544, located in the Taurus molecular cloud complex at a distance of 170 pc \citep{Galli2019}, is one of the most well-studied prestellar cores.
Figure \ref{fig:Herschelmaps} (upper panel) shows L1544  as traced by the Herschel continuum map at 350~$\mu$m.
Its dense ($\sim$10$^{6}$ cm$^{-3}$) and extremely cold (down to $\sim$7 K) central region ($\sim$1000 au) exhibits significant CO depletion \citep{Caselli1999,Crapsi2007, Caselli2022}. In contrast, the outer regions host various chemical processes that produce iCOMs and carbon chain species, as evidenced by single-dish and interferometric observations \citep{Bizzocchi2014,  Vastel2014, Vastel2016, Jimenez2016,  Spezzano2017, Punanova2018,Vastel2018, 2018Hilyblant, Vastel2019,Urso2019}.
More specifically, methanol peak is about 4000 au north of the core center while emission from c-C$_3$H$_2$ and cyanopolyynes peaks in the southern part of the core \citep{Bizzocchi2014, Spezzano2016b, Lin2022, Bianchi2023}. 
This chemical differentiation has been suggested to arise from uneven illumination, influenced by both the asymmetrical 3D structure of the core and its location at the edge of the molecular cloud \citep{Jensen2023}.


\subsection{The IRAS 16293--2422 protostellar system}

IRAS 16293--2422 (hereafter IRAS 16293)
is an archetypical Solar-type star forming region located in the $\rho$ Ophiuchus cloud complex, at a distance $d$ = 141 pc \citep{dzib2018}. IRAS 16293 is the first star forming region where a Class 0 protostar has been identified \citep{Andre1993}, and 
it has been the target of several studies at both mm- and sub-mm wavelengths that have revealed its physical structure (see the Herschel continuum map at 350~$\mu$m in Fig. \ref{fig:Herschelmaps} (Lower panel). 
IRAS 16293 is associated with a large envelope extending up to 6000 au \citep{Castets2001, Crimier2010}, which surrounds a close binary (protostars A1 and A2) separated by 50 au \citep{Maureira2020}, and a third protostellar object (labelled B) located 600 au \citep{Mundy1992, Jorgensen2016} from the A1+A2 system.
Within the same molecular cloud lies the prestellar core IRAS 16293 E, located about 90$\arcsec$ east of the protostellar objects \citep{Loinard2001}.
The inner regions of IRAS 16293 A and B are characterized by the presence of hot corinos, where the gas phase abundances of many molecules, particularly iCOMs, increase by several orders of magnitude compared to those in the surrounding envelope.
IRAS 16293 has been the preferential target of several unbiased spectral line surveys dedicated to unveiling the chemical composition of star-forming regions, such as CHESS \citep{Ceccarelli2010}, TIMASS \citep{Caux2011}, and PILS \citep{Jorgensen2016}. These studies have led to the discovery of a large chemical complexity in the hot corinos around IRAS 16293 A and B \citep[e.g.][and references therein]{Cazaux2003, Bottinelli2004, Jaber2014, Jorgensen2016, Jorgensen2018, Manigand2021, Muller2023}. 
On the other hand, the outer envelope is cold ($\sim$10--30 K), and characterized by the presence of small cyanopolyynes \citep [up to HC$_{5}$N,][]{Jaber2017}, and small carbon chains such as c-C$_{3}$H$_{2}$, C$_{2}$H, C$_{3}$H, C$_{4}$H \citep{Caux2011}, primary revealed thanks to single-dish observations.


\section{Observations} \label{sec:obs}

The observations were conducted on the GBT between June 2019 and February 2020, under project codes AGBT19A\_048 and AGBT20A\_135 (PI: C. Codella). The target sources, L1544 and IRAS 16293–2422, were observed at coordinates $\alpha$$_{\rm J2000}$ = 05$^{h}$ 04$^{m}$ 16$\fs$6 and $\delta$$_{\rm J2000}$ = +25$\degr$ 10$\arcmin$ 48$\farcs$0, and
$\alpha$$_{\rm J2000}$ = 16$^{h}$ 32$^{m}$ 22$\fs$72 and $\delta$$_{\rm J2000}$ = -24$\degr$ 28$\arcmin$ 34$\farcs$3, respectively. 
Study of the cyanopolyynes towards L1544 has  already been presented in \citet{Bianchi2023}.
L1544 and IRAS 16293 were both observed with the Ku band receiver in combination with the VEGAS spectrometer in multiple setups to cover the frequency range (13.5 - 15.4 GHz). The resulting spectral windows had a bandwidth of 187.5 MHz and comprised 131,072 channels, yielding an extremely high spectral resolution of 1.4 kHz (equivalent to $\sim$50 m s$^{-1}$ and 30 m s$^{-1}$ at 9 and 14 GHz, respectively). 
The spectral resolution of the IRAS 16293 spectrum was downgraded to 2.98 kHz for c-C$_3$H, C$_6$H, and HC$_9$N, while a resolution of 24 kHz was used for HC$_7$N to increase the S/N and identify weak lines.
L1544 and IRAS16293 were observed for 6 and 8 hours, respectively. L1544 was successively observed in X band for additional 19 hours to cover the frequency range (8.0-11.6 GHz).  The telescope’s half-power beam width (HPBW) ranged from approximately 54$\arcsec$ in the Ku band to 84$\arcsec$ in the X band. 
In L1544, the HNCO, methanol and c-C$_3$H$_2$ peaks, as reported in \citet{Lin2022}, are included within the largest X beam.
Pointing and focus were set using as calibrators the sources 1733-1304 and 0530+1331 for IRAS 16293 and L1544, respectively. The observations were carried out using position switching mode, employing an ON-OFF throw position of 1$\degr$ west. Data reduction was performed using the GBTIDL software\footnote[4]{https://gbtidl.nrao.edu/}.
 Initially, bad scans were identified and flagged, followed by the performance of ON-OFF position switching subtraction. The calibrated scans from each observing session were then noise-weighted averaged. Subsequently, any RFI (Radio Frequency Interference) and artifacts introduced by Doppler correction were identified and removed. A polynomial fit was applied to remove the baseline. Absolute flux calibration was then performed based on our flux calibrator observations, resulting in a calibration uncertainty of 20\%. The typical root mean square (rms) noise level is $\sim$ 5 mK per channel for both the Ku and X spectra. Line analysis was performed using the GILDAS\footnote[5]{http://www.iram.fr/IRAMFR/GILDAS} CLASS package.
Line intensities were converted to main beam temperature (T$_{\mathrm{MB}}$) using the beam efficiencies of 0.9864 (10 GHz) and 0.959 (15 GHz), and forward efficiency >0.98, reported in the GBT website$^1$.
 


\section{Results} \label{sec:results}
The spectra observed in Ku band for both L1544 and IRAS 16293, and that observed in X band towards L1544, are shown in Fig. \ref{fig:full-band-X} and \ref{fig:full-bands-Ku}, respectively.
The present observations have revealed a significant number of complex carbon chain species, including C$_2$S, C$_3$S, C$_3$N, c-C$_3$H, C$_4$H, C$_6$H, HC$_3$N, HC$_5$N, HC$_7$N, and HC$_9$N. The lines are considered detected when the line peak intensity signal to noise (S/N) is at least 3. Table \ref{Tab:lines} reports the frequency of each transition (MHz), the energies of the upper level E$_{\rm up}$ (K), the line strength S$\mu^2$ (D$^2$), the peak intensity T$_{\rm peak}$ in T$_{\rm MB}$ scale (mK) and the velocity integrated line intensity I$_{\rm int}$ (mK km s$^{-1}$). For L1544, we also include in Table \ref{Tab:lines} the cyanopolyynes lines previously published by \citet{Bianchi2023} from the same dataset.
The most striking result from the Ku-band comparison (13.9–15.41 GHz), where both targets have been observed, is that L1544 exhibits a richer spectrum than the IRAS 16293 envelope (see Fig. \ref{fig:full-band-X}).
More specifically, in IRAS 16293 the only detected molecular species are c-C$_3$H and, in addition, ortho- and para- formaldehyde (H$_2$CO). Conversely, in L1544 we detected in Ku band, formaldehyde, c-C$_3$H, C$_6$H (with associated hyperfine structure), HC$_7$N, and HC$_9$N.  
In addition, the observations of L1544 performed in X band led to the detection of C$_2$S, C$_3$S, C$_3$N, C$_4$H, HC$_5$N, HC$_5$N and further lines of C$_6$H.

\setlength{\tabcolsep}{7.5pt}
\begin{table*}
    \begin{center}
    \label{tab:detections}
\caption{Observed molecular lines towards L1544 and IRS16293. Line peak temperatures and velocity integrated intensities are in T$_{\rm MB}$ scale. \label{Tab:lines}}
 \begin{tabular}{lcrcc||cc||cc}
 \hline
 \hline
 {Species} &  {Transition} &  {$\nu$$^{\rm a}$} &  {E$_{\rm up}$$^{\rm a}$} &  {S$\mu^2$$^{\rm a}$} &  {T$_{\rm peak}$$^{\rm b}$} &  {I$_{\rm int}$$^{\rm b}$} &  {T$_{\rm peak}$$^{\rm b}$} &  {I$_{\rm int}$$^{\rm b}$}\\
 {} &  {} &  {(MHz)} &  {(K)} &  {(D$^2$)} &   {(mK)} &  {(mK km s$^{-1}$)} &  {(mK)} &  {(mK km s$^{-1}$)}\\
 {} &  {} &  {} & {} &  {} & \multicolumn{2}{c}{L1544} & \multicolumn{2}{c}{IRAS 16293$^{\rm c}$}\\
\hline
C$_2$S & N = 2--1, J=1--0 & 11119.446400(6E-7) &  0.53  & 8.3 & 566$\pm$7 & 154$\pm$2 & ... & ... \\
\hline
C$_{\rm 3}$S & J = 2--1 & 11561.513200(9E-7) &  0.83  & 25.9 &  177$\pm$7 & 45$\pm$2 & ... & ...  \\
\hline
C$_{\rm 3}$N & N = 1--0 J=3/2--1/2 F = 3/2-1/2 & 9884.29320(1.6E-5) &  0.47  & 5.6 & 16$\pm$6 & 4$\pm$1 &  ... & ... \\
C$_{\rm 3}$N & N = 1--0 J=3/2--1/2 F = 5/2-3/2 & 9885.89000(1.0E-5) &  0.47  & 16.2 & 44$\pm$7 & 10$\pm$1 &  ... & ... \\
C$_{\rm 3}$N & N = 1--0 J=3/2--1/2 F = 3/2-3/2 & 9886.09420(2.7E-5) &  0.47  & 5.2 & $\leq$ 15 & $\leq$ 3 &  ... & ... \\
C$_{\rm 3}$N & N = 1--0 J=3/2--1/2 F = 1/2-1/2 & 9886.99370(3.0E-5) &  0.47  & 5.0 & $\leq$ 18 & $\leq$ 3 &  ... & ... \\
C$_{\rm 3}$N & N = 1--0 J=3/2--1/2 F = 1/2-3/2 & 9888.79480(3.5E-5) &  0.47  & 0.4 & $\leq$ 21 & $\leq$ 4 &  ... & ... \\
C$_{\rm 3}$N & N = 1--0 J=1/2--1/2 F = 1/2-1/2 & 9911.18670(4.0E-5) &  0.47  & 0.5 & $\leq$ 18 & $\leq$ 3 &  ... & ... \\
C$_{\rm 3}$N & N = 1--0 J=1/2--1/2 F = 1/2-3/2 & 9912.98770(3.6E-5) &  0.47  & 5.0 & $\leq$ 18 & $\leq$ 3 &  ... & ... \\
C$_{\rm 3}$N & N = 1--0 J=1/2--1/2 F = 3/2-1/2 & 9913.64600(3.0E-5) &  0.47  & 5.2 & $\leq$ 21 & $\leq$ 4 &  ... & ... \\
C$_{\rm 3}$N & N = 1--0 J=1/2--1/2 F = 3/2-3/2 & 9915.44700(1.6E-5) &  0.47  & 5.6 & $\leq$ 21 & $\leq$ 4 &  ... & ... \\
\hline
c-C$_{\rm 3}$H & N = 1$_{\rm 1,0}$--1$_{\rm 1,1}$ J=1/2--1/2 F = 1-1 & 14686.63000(4.0E-6) &  0.71  & 3.8 &  71$\pm$7 & 23$\pm$1 &  21$\pm$4 & 13$\pm$2 \\
c-C$_{\rm 3}$H & N = 1$_{\rm 1,0}$--1$_{\rm 1,1}$ J=1/2--1/2 F = 0-1 & 14689.71800(1.2E-5) &  0.71  & 1.4 &  36$\pm$6 & 7$\pm$1 & $\leq$ 24 & $\leq$ 5 \\
c-C$_{\rm 3}$H & N = 1$_{\rm 1,0}$--1$_{\rm 1,1}$ J=1/2--1/2 F = 1-0 & 14697.69200(8.0E-6) &  0.71  & 2.6 &  46$\pm$7 & 14$\pm$2 &  $\leq$ 24 & $\leq$ 5 \\
c-C$_{\rm 3}$H & N = 1$_{\rm 1,0}$--1$_{\rm 1,1}$ J=1/2--3/2 F = 0-1 & 14755.39620(1.2E-5) &  0.71  & 1.6 &  34$\pm$8 & 11$\pm$2 &  $\leq$ 21 & $\leq$ 5 \\
c-C$_{\rm 3}$H & N = 1$_{\rm 1,0}$--1$_{\rm 1,1}$ J=1/2--3/2 F = 1-2 & 14767.70000(8.0E-6) &  0.71  & 2.8 &  66$\pm$7 & 15$\pm$2 &  15$\pm$5 & 11$\pm$1 \\
c-C$_{\rm 3}$H & N = 1$_{\rm 1,0}$--1$_{\rm 1,1}$ J=1/2--3/2 F = 2-1 & 14812.01000(8.0E-6) &  0.71  & 1.7 & 43$\pm$7 & 11$\pm$2 & $\leq$ 24 & $\leq$ 5 \\
c-C$_{\rm 3}$H & N = 1$_{\rm 1,0}$--1$_{\rm 1,1}$ J=1/2--3/2 F = 1-1 & 14829.57100(6.4E-6) &  0.71  & 2.0 & 43$\pm$8 &  12$\pm$2 & $\leq$ 24 & $\leq$ 5 \\
c-C$_{\rm 3}$H & N = 1$_{\rm 1,0}$--1$_{\rm 1,1}$ J=1/2--3/2 F = 1-0 & 14(1.0E-5) &  0.71  & 0.7 & $\leq$ 24 & $\leq$ 5 & $\leq$ 24 & $\leq$ 5 \\
c-C$_{\rm 3}$H & N = 1$_{\rm 1,0}$--1$_{\rm 1,1}$ J=1/2--3/2 F = 2-1 & 14877.67500(4.0E-6) &  0.71  & 2.0 & 41$\pm$6 & 12$\pm$2 & $\leq$ 24 & $\leq$ 5 \\
c-C$_{\rm 3}$H & N = 1$_{\rm 1,0}$--1$_{\rm 1,1}$ J=3/2--3/2 F = 2-2 & 14893.05100(4.0E-6) &  0.71  & 11.1 & 211$\pm$5 & 71$\pm$1 & 58$\pm$9 & 40$\pm$2 \\
c-C$_{\rm 3}$H & N = 1$_{\rm 1,0}$--1$_{\rm 1,1}$ J=3/2--3/2 F = 1-1 & 14895.24300(8.0E-6) &  0.72  & 5.3 & 104$\pm$5 & 71$\pm$1 & 27$\pm$3 & 14$\pm$2 \\
c-C$_{\rm 3}$H & N = 1$_{\rm 1,0}$--1$_{\rm 1,1}$ J=3/2--3/2 F = 1-2 & 14910.62500(4.0E-6) &  0.72  & 0.9 & 22$\pm$6 & 5$\pm$1 & $\leq$ 21 & $\leq$ 5 \\
\hline
C$_{\rm 4}$H & N = 1--0 J=3/2--1/2 F = 1--0 & 9493.06000(2.0E-6) & 0.46 & 2.5 & 86$\pm$7 & 19$\pm$2 & ... & ... \\
C$_{\rm 4}$H & N = 1--0 J=3/2--1/2 F = 2--1 & 9497.61500(2.0E-6) & 0.46 & 7.4 & 247$\pm$6 & 59$\pm$1 & ... & ... \\
C$_{\rm 4}$H & N = 1--0 J=3/2--1/2 F = 1--1 & 9508.00500(2.0E-6) & 0.46 & 1.9 & 69$\pm$7 & 16$\pm$2 & ... & ... \\
C$_{\rm 4}$H & N = 1--0 J=1/2--1/2 F = 1--0 & 9547.96000(2.0E-6) & 0.46 & 1.9 & 70$\pm$6 & 20$\pm$1 & ... & ... \\
C$_{\rm 4}$H & N = 1--0 J=1/2--1/2 F = 0--1 & 9551.72000(2.0E-6) & 0.46 & 1.5 & 68$\pm$7 & 15$\pm$1 & ... & ... \\
C$_{\rm 4}$H & N = 1--0 J=1/2--1/2 F = 1--1 & 9562.90500(2.0E-6) & 0.46 & 2.5 & 93$\pm$7 & 21$\pm$2 & ... & ... \\
\hline
C$_{\rm 6}$H & J=7/2--5/2 $\Omega$=3/2 F=4--3 l=e & 9703.5080(1.0E-2) & 0.80 & 111.9 & 41$\pm$6 & 7$\pm$2 &  ... & ... \\
C$_{\rm 6}$H & J=7/2--5/2 $\Omega$=3/2 F=3--2 l=e & 9703.6000(1.0E-2) & 0.80 & 82.9 & 28$\pm$6 & 8$\pm$2 &  ... & ... \\
C$_{\rm 6}$H & J=7/2--5/2 $\Omega$=3/2 F=4--3 l=f & 9703.8350(1.0E-2) & 0.80 & 111.9 & 46$\pm$6 & 10$\pm$2 &  ... & ... \\
C$_{\rm 6}$H & J=7/2--5/2 $\Omega$=3/2 F=3--2 l=f & 9703.9360(1.0E-2) & 0.80 & 82.9 & 31$\pm$6 &  6$\pm$2 &  ... & ... \\
C$_{\rm 6}$H & J=7/2--5/2 $\Omega$=1/2 F=4--3 l=f & 9758.8270(1.0E-2) & 22.68 & 134.3 & $\leq$ 18 & $\leq$ 3 &  ... & ... \\
C$_{\rm 6}$H & J=7/2--5/2 $\Omega$=1/2 F=3--2 l=f & 9759.1166(9.2E-3) & 22.68 & 99.5 & $\leq$ 18 & $\leq$ 3 &  ... & ... \\
C$_{\rm 6}$H & J=7/2--5/2 $\Omega$=1/2 F=4--3 l=e & 9786.7810(4.0E-3) & 22.68 & 134.2 & $\leq$ 21 & $\leq$ 4 &  ... & ... \\
C$_{\rm 6}$H & J=7/2--5/2 $\Omega$=1/2 F=3--2 l=e & 9787.0050(4.0E-3) & 22.68 & 99.5 & $\leq$ 18 & $\leq$ 3 &  ... & ... \\
C$_{\rm 6}$H & J=11/2--9/2 $\Omega$=3/2 F=6--5 l=e & 15248.2470(5.0E-3) & 2.12 & 192.0 & 117$\pm$5 & 23$\pm$1 & $\leq$ 18 & $\leq$ 3 \\
C$_{\rm 6}$H & J=11/2--9/2 $\Omega$=3/2 F=5--4 l=e & 15248.3320(5.0E-3) & 2.12 & 159.5 & 107$\pm$5 & 19$\pm$1 & $\leq$ 18 & $\leq$ 3 \\
C$_{\rm 6}$H & J=11/2--9/2 $\Omega$=3/2 F=6--5 l=f & 15249.0840(5.0E-3) & 2.12 & 192.0 & 113$\pm$5 & 19$\pm$1 &  $\leq$ 18 & $\leq$ 3 \\
C$_{\rm 6}$H & J=11/2--9/2 $\Omega$=3/2 F=5--4 l=f & 15249.1580(5.0E-3) & 2.12 & 159.5 & 100$\pm$5 & 20$\pm$1 & $\leq$ 18 & $\leq$ 3 \\
C$_{\rm 6}$H & J=11/2--9/2 $\Omega$=1/2 F=6--5 l=f & 15343.1670(5.0E-3) & 24.02 & 205.7 & $\leq$ 18 & $\leq$ 3 & $\leq$ 18 & $\leq$ 3 \\
C$_{\rm 6}$H & J=11/2--9/2 $\Omega$=1/2 F=5--4 l=f & 15343.2370(5.0E-3) & 24.02 & 170.9 & $\leq$ 18 & $\leq$ 3 & $\leq$ 18 & $\leq$ 3 \\
C$_{\rm 6}$H & J=11/2--9/2 $\Omega$=1/2 F=6--5 l=e & 15371.5050(5.0E-3) & 24.02 & 205.7 & $\leq$ 18 & $\leq$ 3 & $\leq$ 18 & $\leq$ 3 \\
C$_{\rm 6}$H & J=11/2--9/2 $\Omega$=1/2 F=5--4 l=e & 15371.5600(5.0E-3) & 24.02 & 170.9 & $\leq$ 18 & $\leq$ 3 & $\leq$ 18 & $\leq$ 3 \\
\hline
HC$_{\rm 7}$N & J=13--12 & 14663.9937(2.0E-4) & 4.93 & 906.1 & 443$\pm$7$^{\rm d}$ & 151$\pm$1$^{\rm d}$ & 9$\pm$2 &  9$\pm$2 \\
\hline
HC$_{\rm 9}$N & J=24--23 & 13944.8310(1.0E-3) & 8.37 & 1947.0 & 152$\pm$6$^{\rm d}$ & 24$\pm$1$^{\rm d}$ & $\leq$ 18 & $\leq$ 2 \\
HC$_{\rm 9}$N & J=25--24 & 14525.8620(1.0E-3) & 9.06 & 2028.1 & 157$\pm$6$^{\rm d}$ & 25$\pm$1$^{\rm d}$ & $\leq$ 15 & $\leq$ 2 \\
HC$_{\rm 9}$N & J=26--25 & 15106.8910(1.0E-3) & 9.79 & 2108.9 & 166$\pm$6$^{\rm d}$ & 26$\pm$1$^{\rm d}$ & $\leq$ 15 & $\leq$ 2 \\
\hline
\end{tabular}
\end{center}
$^{\rm a}$ Frequencies and spectroscopic parameters have been provided by \citet{Yamamoto1990}, \citet{Yamamoto1987}, 
and \citet{Yamamoto1994} for C$_2$S, C$_3$S, and c-C$_3$H, respectively, and extracted from the Jet Propulsion Laboratory molecular database \citep[][and \url{https://spec.jpl.nasa.gov/}]{Pickett1998}.
On the other hand, frequencies and spectroscopic parameters
for C$_3$N \citep{Mikami1989}, C$_4$H \citep{Gottlieb1983}, and C$_6$H \citep{McCarthy1999,Gottlieb2010}, are extracted from  the Cologne Database for Molecular Spectroscopy \citep[][and \url{http://www.astro.uni-koeln.de/cdms/}]{Endres2016, Muller2001, Muller2005}. 
$^{\rm b}$ Errors do not include 20$\%$ of calibration. 
Uncertainties on T$_{\rm peak}$ are the local rms of the spectra. Upper limits refer to the 3$\sigma$ level. Line integrated intensities (I$_{\rm int}$) are computed by carefully selecting and integrating the emission over the relevant velocity range.
$^{\rm c}$ '...' means that the frequency of the transition is not in the observed frequency range. 
$^{\rm d}$ Data presented in \cite{Bianchi2023}.
\end{table*}

\begin{figure}
\begin{center}
\includegraphics[scale=0.55]{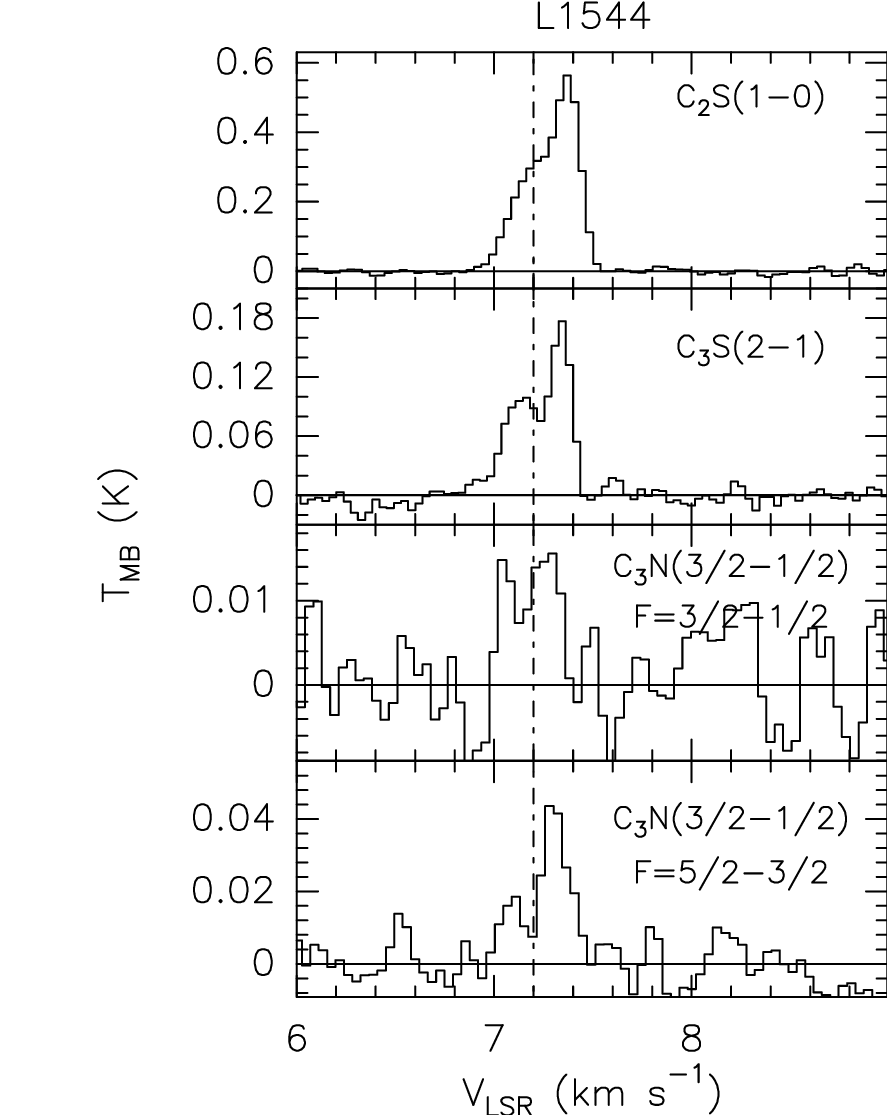}
  \caption{C$_2$S, C$_{\rm 3}$S, and C$_{\rm 3}$N spectra (in T$_{\rm MB}$ scale) observed towards L1544. The vertical dashed lines mark the ambient LSR velocity (+7.2 km s$^{-1}$, \citealt{Tafalla1998}).} 
  \label{fig:spectra-CCS}
  \end{center}
\end{figure}

\begin{figure*}
\includegraphics[scale=0.65]{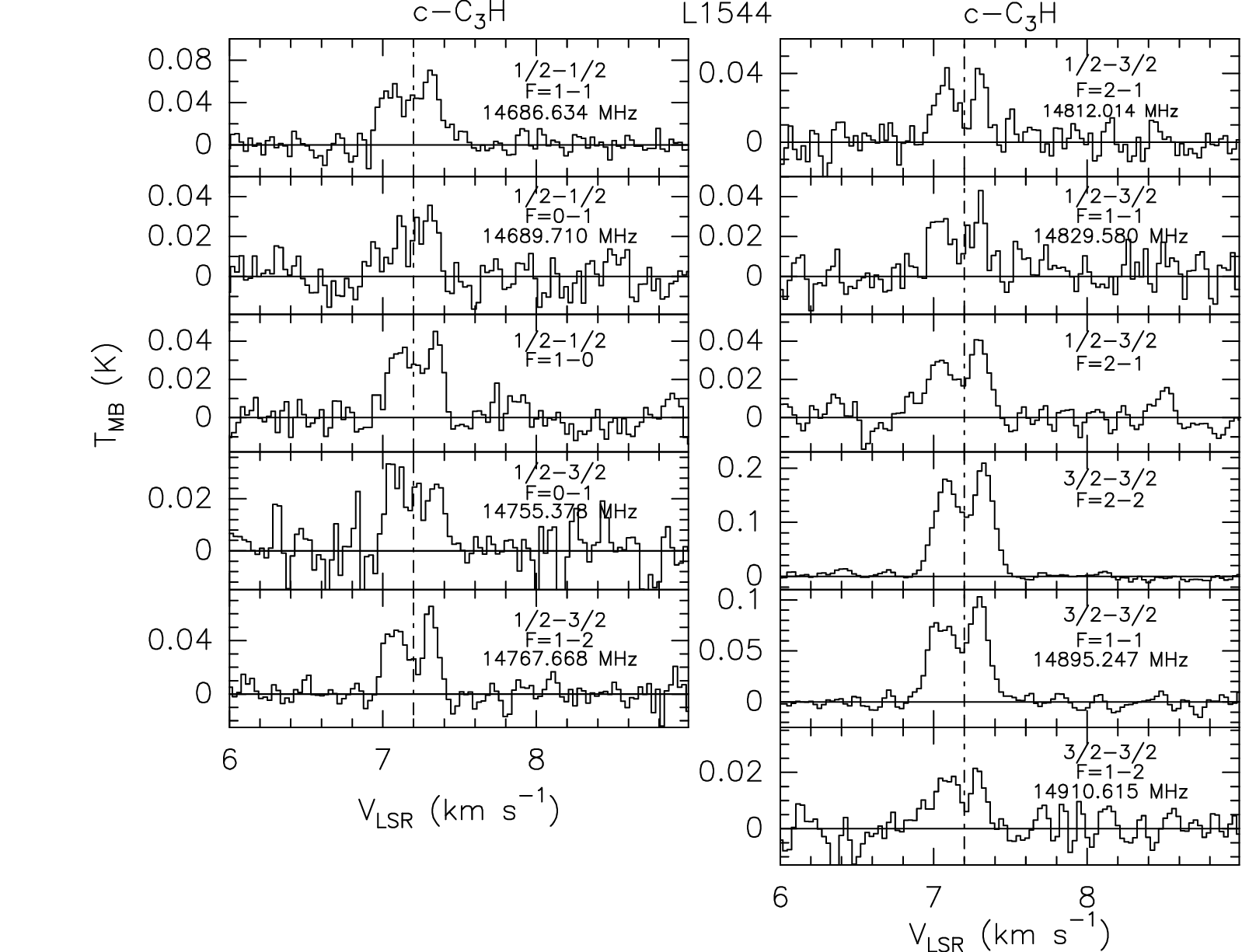}
  \caption{c--C$_3$H spectra (in T$_{\rm MB}$ scale) observed towards L1544. The vertical dashed lines mark the ambient LSR velocity (+7.2 km s$^{-1}$, \citealt{Tafalla1998}). When reported, the frequencies (in MHz) are different
  from those extracted from the JPL catalogue \citep{Pickett1998}, and refer to the values needed to center the spectra to the L1544 systemic velocity.}
  \label{fig:spectra-C3H-L1544}
\end{figure*}

\begin{figure*}
\includegraphics[scale=0.65]{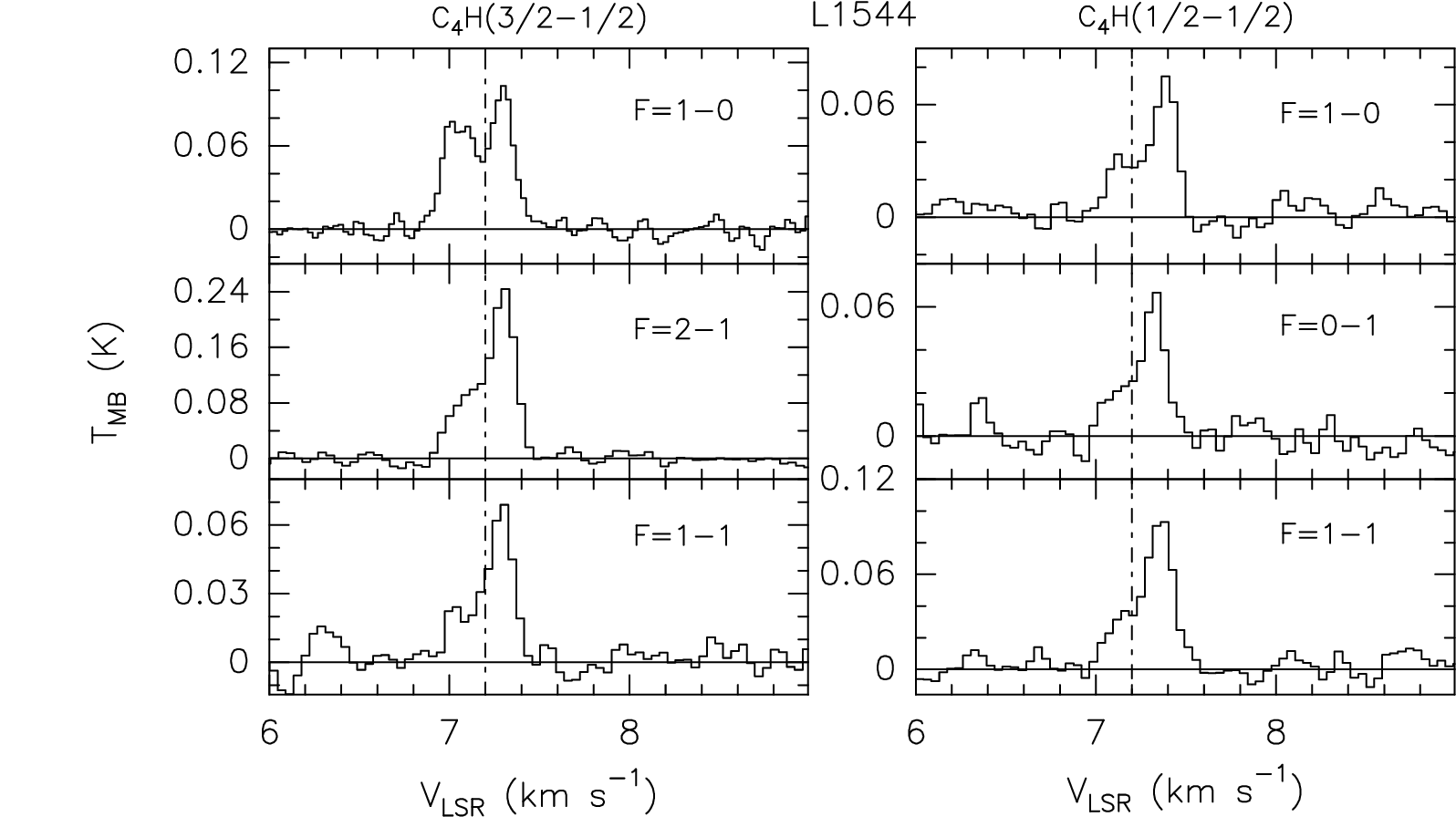}
  \caption{C$_4$H spectra (in T$_{\rm MB}$ scale) observed towards L1544. The vertical dashed lines mark the ambient LSR velocity (+7.2 km s$^{-1}$, \citealt{Tafalla1998}).}
  \label{fig:spectra-C4H}
\end{figure*}

\begin{figure*}
\includegraphics[scale=0.65]{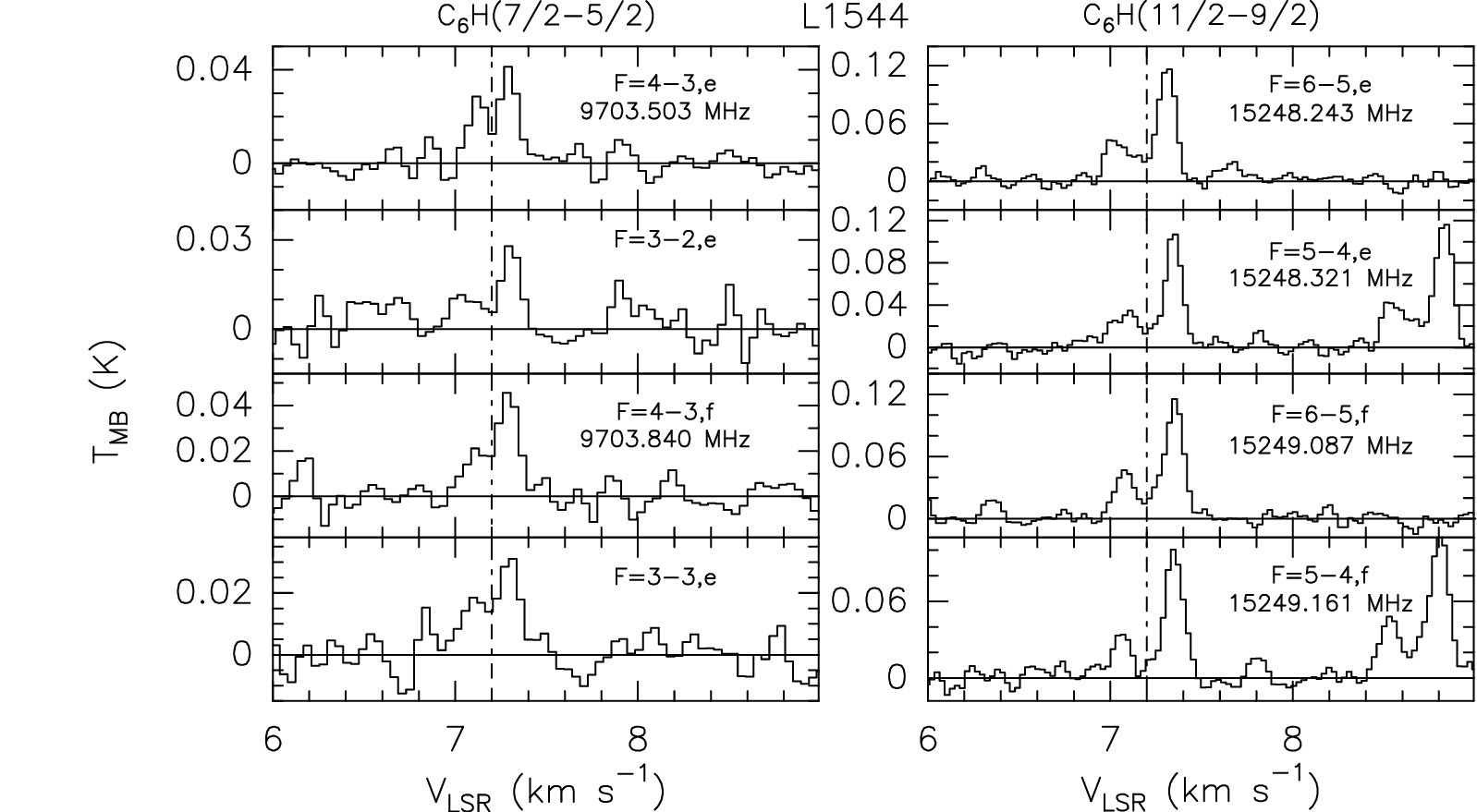}
  \caption{C$_6$H spectra (in T$_{\rm MB}$ scale) observed towards L1544. The vertical dashed lines mark the ambient LSR velocity (+7.2 km s$^{-1}$, \citealt{Tafalla1998}). When reported, the frequencies (in MHz) are different
  from those extracted from the CDMS catalogue \citep{Muller2005}, and refer to the values needed to center the spectra to the L1544 systemic velocity.}
  \label{fig:spectra-C6H}
\end{figure*}

\begin{figure}
\includegraphics[scale=0.65]{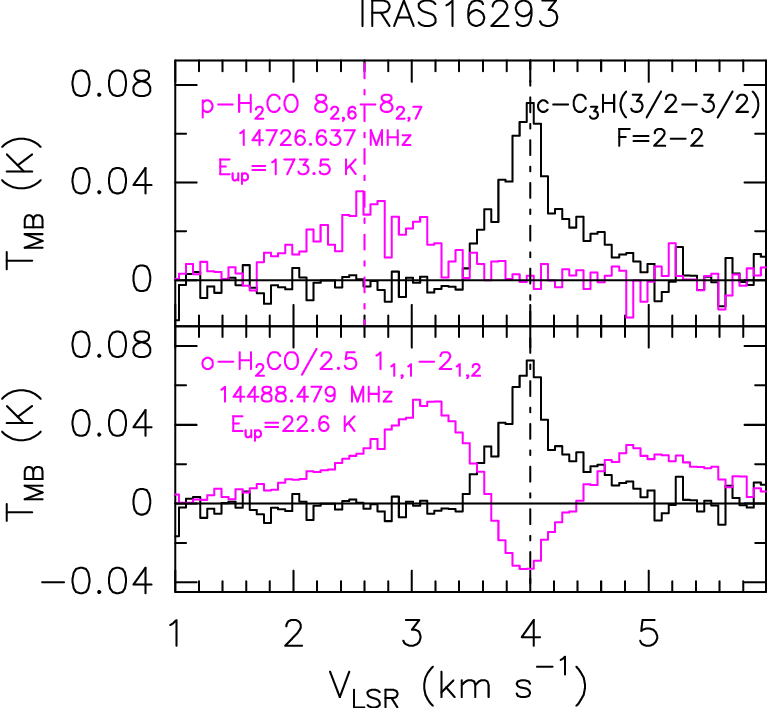}
  \caption{Upper panel: Comparison between the c--C$_3$H(3/2-3/2) F = 2-2  (black) and p-H$_2$CO(8$_{2,6}$--8$_{2,7}$) (E$_{\rm up}$ = 174  K, magenta) line profiles observed towards IRAS 16293. The vertical scale is in T$_{\rm MB}$ units. The vertical dashed black line marks the LSR velocity of the envelope \citep[+4.0 km s$^{-1}$,][]{Caux2011}, while the magenta one is for the LSR velocity of the hot gas around the A and B protostars \citep[+2.6 km s$^{-1}$,][]{Caux2011}. Lower panel: Same as the upper panel using the low-excitation (E$_{\rm up}$ = 23 K) o-H$_2$CO(1$_{1,1}$--2$_{1,2}$) emission line (divided by a factor 2.5 for sake of clarity).}
  \label{fig:spectra-C3H-H2CO}
\end{figure}

\begin{figure}
\includegraphics[scale=0.65]{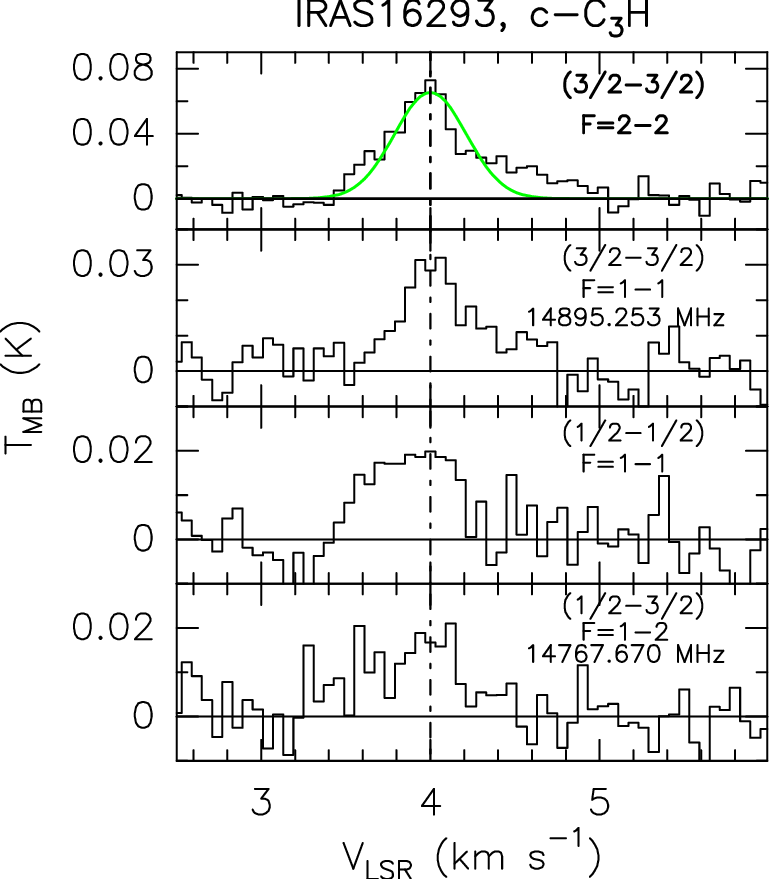}
  \caption{c--C$_3$H spectra (in T$_{\rm MB}$ scale) observed towards IRAS 16293. The spectral resolution has been downgraded to 2.98 kHz ($\sim$60 m s$^{-1}$) to improve the S/N ratio.
  The green curve shows the envelope emission (see text).
The vertical dashed lines mark the ambient LSR velocity (+4.0 km s$^{-1}$, \citealt{Caux2011}). When reported, the frequencies (in MHz) are different from those extracted from the JPL catalogue \citep{Pickett1998}, and refer to the values needed to center the spectra to the IRAS 16293 systemic velocity.}
  \label{fig:spectra-C3H}
\end{figure}

\begin{figure}
\includegraphics[scale=0.63]{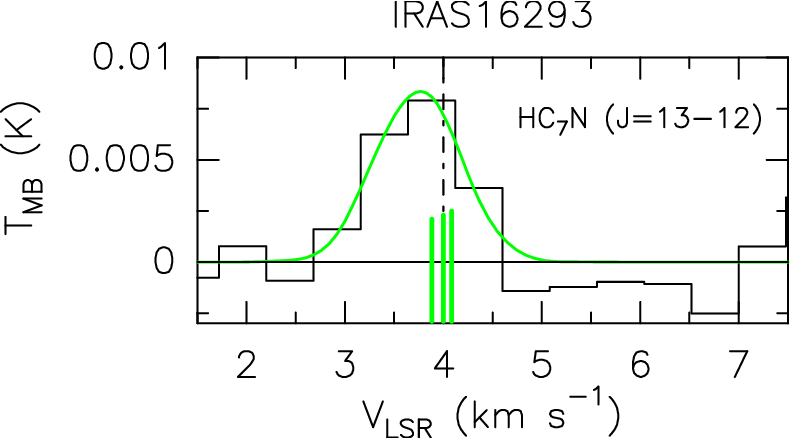}
  \caption{HC$_7$N(13--12) spectra (in T$_{\rm MB}$ scale) detected towards IRAS 16293. The spectral resolution has been downgraded to 24 kHz ($\sim$ 0.5 km s$^{-1}$) to increase the S/N ratio. The vertical dashed line marks the ambient LSR velocity of the envelope \citep[+4.0 km s$^{-1}$,][]{Caux2011}. The green curve represents the Gaussian fit, while the vertical solid segments indicate the brightest hyperfine transitions (F = 12--11, 13--12, and 14--13, in order of increasing rest frequency) expected to contribute to the observed profile.
  }
  \label{fig:spectra-HC7N}
\end{figure}



\subsection{L1544}\label{subsec:L1544-LTE}

In L1544, we detected a single emission line from C$_2$S and C$_3$S, two lines from C$_3$N, eleven lines from c-C$_3$H, six lines from C$_4$H, and eight lines from C$_6$H (see Tab.~\ref{Tab:lines}). In addition, the HC$_{\rm 3}$N, HC$_{\rm 5}$N, HC$_{\rm 7}$N, and HC$_{\rm 9}$N lines detected in L1544 in the present dataset have been analysed in the study by \citet{Bianchi2023}.
The spectra of all the detected transitions for each species are shown in Figures~\ref{fig:spectra-CCS}, \ref{fig:spectra-C3H-L1544}, \ref{fig:spectra-C4H} and \ref{fig:spectra-C6H}. The exceptionally high spectral resolution of the present dataset ($\sim$50 m s$^{-1}$ and 30 m s$^{-1}$ at 9 and 14 GHz, respectively; see Sect. \ref{sec:obs}) enables excellent sampling of the line profiles.  All the detected species exhibit a double-peaked line profile, blue- and red-shifted by 0.1--0.2 km s$^{-1}$ with respect to the L1544 systemic velocity \citep[+7.2 km s$^{-1}$, e.g.][]{Tafalla1998}. The red-shifted peak has a higher intensity than the blue-shifted one. 
Moreover, the spectral profiles of the current dataset closely align with those previously observed in GBT K-Band toward L1544 for C$_4$H and C$_6$H, as reported by \citet{Gupta2009}.
This spectral characteristic profile has also been observed for cyanopolyynes \citep{Bianchi2023}, supporting the idea that these molecules trace the same gas component. 
Given the asymmetric line profiles, we calculated the line-integrated intensities by carefully selecting and integrating the emission over the relevant velocity range.
For non-detections, a 3$\sigma$ upper limit on the integrated intensity (I$_{\rm int}$) was estimated as $3\sqrt{N_{\rm ch}}$$d$v$\sigma$, where $N_{\rm ch}$ is the number of channels included in the integration (corresponding to a line width of 0.4 km s$^{-1}$, \citealt{Bianchi2023}) and $d$v is the velocity resolution. We included a calibration uncertainty of 20\% in the calculation. 
To estimate the total column densities, $N^{\rm tot}$, of the detected species, we used the integrated line intensities reported in Tab.~\ref{Tab:lines}.
Given the absence of available collisional coefficients for c-C$_3$H and C$_4$H, and the detection of only a single line for C$_2$S, C$_3$S, C$_3$N and C$_6$H, we performed a Local Thermodynamic Equilibrium (LTE) analysis for these species.
We assume that the emission fills the beam and we verify this assumption for C$_6$H, the only species having multiple detected transitions both in X- and Ku-band. In this case, the column densities derived using different transitions in X- and Ku-band independently are consistent within the uncertainties. We assume that the same is valid also for the other species. We derived column densities assuming excitation temperatures of both 5 K and 12 K, and we report the results as ranges of values. These limits are consistent with the excitation temperatures found by \citet{Bianchi2023} for the cyanopolyynes using the Large Velocity Gradient (LVG) method.
Consequently, throughout this work we assume that the emission is optically thin and in LTE at the temperatures derived by \citet{Bianchi2023}. 
For c-C$_3$H, C$_4$H and C$_6$H we used the hyperfine fitting method available in GILDAS to fit the hyperfine components to the available astronomical data. We confirm that emission is optically thin ($\tau$<0.1) for all species, in agreement with what assumed in the present analysis. 
For C$_2$S, C$_3$S, C$_3$N and C$_6$H we calculate the critical density of the detected transition using the Einstein and collisional coefficients from the LAMDA database\footnote[6]{https://home.strw.leidenuniv.nl/~moldata/} \citep{Schoier2005}, and from \citet{walker2018}, \citet{Sahnoun2020} and \citet{laramoreno2021}. 
For all species, we estimate n$_{\rm crit}$ $\lesssim$ 8 $\times$ 10$^3$ cm$^{-3}$ at 10 K and n$_{\rm crit}$ $\lesssim$ 1 $\times$ 10$^4$ cm$^{-3}$ at 20 K.
In order to verify if the LTE assumption is valid, we consider the physical structure of the two sources.
For L1544, the temperature is <12~K and the gas density is $\gtrsim$ 2 $\times$ 10$^4$ cm$^{-3}$ within a radius of 7140 au probed by the GBT observations in X band, following the density profile by \citet{Keto2010}.
The same holds for IRAS 16293, where the density profile by \citet{Crimier2010} predicts a temperature of >15~K and gas densities $\gtrsim$ 4 $\times$ 10$^5$ cm$^{-3}$ within a radius of 3807 au probed by the GBT observations in Ku band. 
For IRAS 16293 we also checked the critical densities for temperature higher than 20 K, which correspond to the inner regions. For T>20 K, the gas density is larger than 1 $\times$ 10$^6$ cm$^{-3}$, while critical densities are always lower than 5 $\times$ 10$^4$ cm$^{-3}$.
The LTE assumption is thus verified for the transitions of C$_2$S, C$_3$S, C$_3$N and C$_6$H. However, for other transitions from c-C$_3$H and C$_4$H, collisional coefficients are not available. If the lines are sub-thermally populated, the column densities derived under the LTE assumption should be considered as lower limits.

\subsubsection{The {\rm C$_n$X} family: {\rm C$_2$S, C$_3$S, C$_3$N} and {\rm c-C$_3$H}}

We detected C$_2$S (N = 2--1, J = 1--0) and C$_3$S(2--1)
(see Fig.~\ref{fig:spectra-CCS}), toward L1544.
When considering the velocity-integrated emission and a temperature range between 5 and 12 K, we obtain $N^{\rm tot}_{\rm C_2S}$ = 1.1--2.0 $\times$ 10$^{13}$ cm$^{-2}$,
and $N^{\rm tot}_{\rm C_3S}$ = 1.5-2.0 $\times$ 10$^{12}$ cm$^{-2}$. 
On the other hand, the C$_3$N (3/2--1/2) line (assuming again temperature in the 5--12 K range) leads to 
$N^{\rm tot}_{\rm C_3N}$ = 2--4 $\times$ 10$^{12}$ cm$^{-2}$.
In all these cases, the column densities derived from the red-shifted peaks are $\sim$ 2-3 times higher than those derived from the blue-shifted ones, consistent with the fact that the emission mainly originates from the southern region of L1544, as found for cyanopolyynes by \citet{Spezzano2017} and \citet{Bianchi2023}. 
The derived beam-averaged column densities are reported in Tab.~\ref{Tab:column-densities}. 
These values are consistent within a factor 3 with previous measurements obtained using the IRAM 30-m at 3 mm for C$_2$S, C$_3$S \citep{2018vastel-Sbearing}, and C$_3$N \citep{Vastel2019}.

For c-C$_3$H, we detect eleven hyperfine components corresponding to the N=1$_{1,0}$–1$_{1,1}$ transition (see Fig.~\ref{fig:spectra-C3H-L1544}), allowing us to derive $N^{\rm tot}_{\rm C_3H}$ between 2 $\times$ 10$^{12}$ cm$^{-2}$ (5 K), and 2 $\times$ 10$^{13}$ cm$^{-2}$ (12 K). 
These values are in agreement with those derived by 
\citep{Mangum1990} at $\sim$ 15 GHz using the GBT 100-m antenna. 

\setlength{\tabcolsep}{3.5pt}
\renewcommand{\arraystretch}{1.2}
\begin{table}
\begin{center}
\caption{Carbon chains column densities and abundances measured towards L1544 and IRAS 16293.
The column densities are estimated assuming a gas temperature between 5 and 12 K in L1544 \citep{Bianchi2023}, and between 5 and 20 K in the envelope of IRAS 16293 \citep{Jaber2017}, respectively. The abundances are derived assuming N(H$_2$) of 2$\times$10$^{22}$ cm$^{-2}$ for L1544 and 5$\times$10$^{22}$ cm$^{-2}$ for IRAS 16293 (based on Herschel observations, see text). 
\label{Tab:column-densities}}
 \begin{tabular}{l|c|c|c|c|c}
 \hline
 \hline
{Species} & \multicolumn{2}{c}{$N_{\rm tot}$ ($\times$10$^{12}$ cm$^{-2}$)} & \multicolumn{2}{c}{Abundances ($\times$10$^{-10}$)} & Reference \\
{} & \multicolumn{1}{c}{L1544} & \multicolumn{1}{c}{IRAS 16293} &  \multicolumn{1}{c}{L1544} & \multicolumn{1}{c}{IRAS 16293}\\
\hline
C$_{\rm 2}$S   & 11--20      & --                & 5.5--10.0 & - & 1 \\
C$_{\rm 3}$S   & 1.5--2.0    & --                & 0.7--1.0  & - & 1 \\
C$_{\rm 3}$N   & 2.0--3.6    & --                & 1.0--1.8  & - & 1 \\
c-C$_{\rm 3}$H & 2.4--17.8   & 1.4--8.9          & 1.2--8.9  & 0.3--1.8 & 1 \\
C$_{\rm 4}$H   & 13.7--30.9  & --                & 6.8--15.5 & - & 1 \\
C$_{\rm 6}$H   & 0.58--1.79  & $\leq$ 0.1--0.5 & 0.3--0.9  & < 0.02--0.1 & 1 \\
HC$_{\rm 3}$N  & > 80        &   --              & > 40      & - & 2 \\
HC$_{\rm 5}$N  & 30--200     & --                & 15--100   & - & 2 \\
HC$_{\rm 7}$N  & 4--50       & 0.06--0.11        & 2--25     & 0.012-0.022 & 1,2 \\
HC$_{\rm 9}$N  & 0.8--16.0   & $\leq$ 0.029--0.032 & 0.4--8.0  & < 0.0058--0.064 & 1,2 \\
\hline
 \end{tabular}
 \end{center}
[1] This work;
[2] \citealt{Bianchi2023}
\\
\end{table}

\subsubsection{The polyynyl radicals family: {\rm C$_4$H} and {\rm C$_6$H}}

We detected C$_4$H thanks to three hyperfine components of the N = 1–0 J=3/2-1/2 spectral pattern, as well as the N = 1–0 J=1/2-1/2 triplet (see Fig.~\ref{fig:spectra-C4H}). C$_6$H is detected in both X and Ku bands (see Fig.~\ref{fig:spectra-C6H}). Specifically, four hyperfine components of the J=7/2–5/2 transition with upper level energy of 0.80 K are detected in X band ($\sim$ 9 GHz), while four hyperfine components of the J=11/2–9/2 transition with upper level energy of 2.1 K are detected in Ku band ($\sim$ 15 GHz).
The line profiles of both C$_4$H, and C$_6$H show the characteristic double-peaked structure observed in L1544 using C$_2$S, C$_3$S, C$_3$N and c-C$_3$H (Sec.~\ref{subsec:L1544-LTE}), and cyanopolyynes \citep{Bianchi2023}, with the red-shifted peak brighter (by a factor $\sim$ 2) than the blue-shifted one.

From the LTE analysis of the velocity-integrated emission, we derived $N^{\rm tot}_{\rm C_4H}$ = 1 $\times$ 10$^{13}$ cm$^{-2}$ and 3 $\times$ 10$^{13}$ cm$^{-2}$ for temperatures of 5 K and 12 K, respectively. 
On the other hand, $N^{\rm tot}_{\rm C_6H}$
is 0.58-1.79 $\times$ 10$^{12}$ cm$^{-2}$, adopting the 5--12 K range. 
These values are well consistent with previous GBT observations by \citet{Gupta2009}.
In contrast, our current measurement of the C$_4$H column density, derived using spectral parameters from the CDMS catalog (see Tab. \ref{Tab:lines}), is an order of magnitude lower than that reported by \citet{nagy2019} and \citet{Gupta2009}, who presumably used the JPL catalog \citep{Pickett1998}. This discrepancy will be discussed in detail in Sect. \ref{sec:discussion}.


\subsection{IRAS 16293}\label{subsec:IRAS-LTE}
\begin{figure}
\includegraphics[scale=0.63]{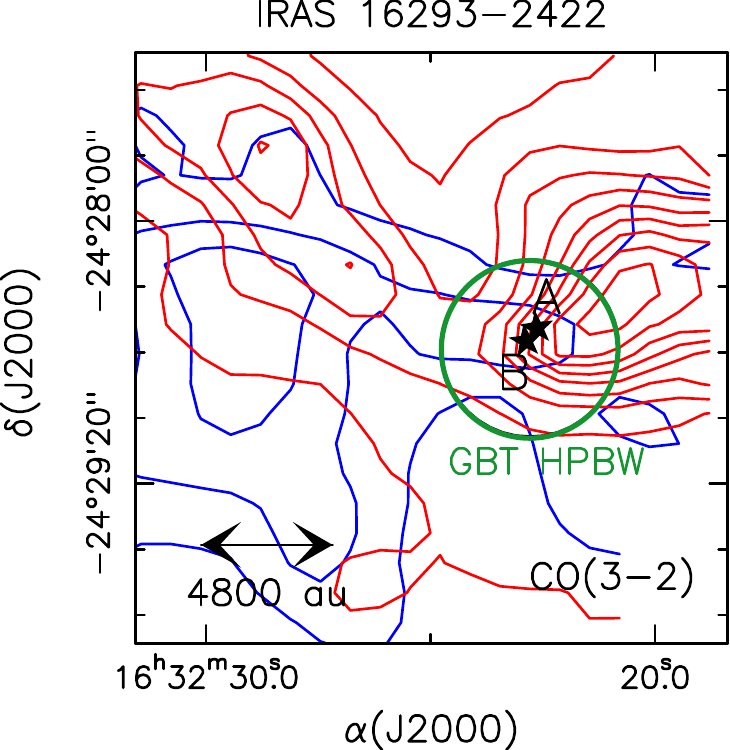}
  \caption{The red- and blue-shifted outflowing activity in the IRAS 16293 region as mapped by CO(3--2) using APEX observations adapted from \citet{Kahle2023}. First steps and contours correspond to 3$\sigma$ (2 K km s$^{-1}$) and 8$\sigma$, respectively. Black stars indicates the positions of the
  IRAS 16293 A and B protostars \citep[e.g.][]{Jorgensen2016}. 
  The green circle indicates the HPBW of the present GBT observations in the X band (54$\arcsec$).}
  \label{fig:IRAS-outflows}
\end{figure}

The observations of IRAS 16293 show the detection of c-C$_3$H, H$_2$CO and HC$_7$N. The transitions of C$_2$S, C$_3$S, C$_3$N, and C$_4$H were not covered by the present observations. On the other hand, we provide upper limits for C$_6$H, and HC$_9$N. The c-C$_3$H line peaks at approximately +4 km s$^{-1}$, consistent with the systemic velocity of the IRAS16293 envelope \citep{Caux2011}. The c-C$_3$H and HC$_7$N transitions are single-peaked, associated with low upper level energies (E$_{\rm up}$ $<$ 25 K) and the line profiles are narrow, with a FWHM $\leq$ 1 km s$^{-1}$. 
These findings support that the emission likely originates from the cold IRAS 16293 envelope, and not from the hot corino hosted inside the cloud. The high excitation (E$_{\rm up}$ = 173.5 K) p-H$_2$CO(8$_{2,6}$--8$_{2,7}$) transition peaks at around +2.6 km s$^{-1}$, the systemic source velocity of the IRAS 16293 B protostar \citep{Caux2011}. Conversely, the low-excitation o-H$_2$CO(1$_{1,1}$--2$_{1,2}$) line (E$_{\rm up}$ = 23 K) exhibits an absorption dip at the velocity of the envelope. 
This suggests absorption of the H$_2$CO(1$_{1,1}$--2$_{1,2}$) emission by the cold gas of the envelope.
Figure~\ref{fig:spectra-C3H-H2CO} (upper panel) presents a comparison between the c-C$_3$H(3/2-3/2) F = 2-2 spectrum and the p-H$_2$CO (8$_{2,6}$--8$_{2,7}$) transition with E$_{\rm up}$ = 174 K. The differing peak velocities of these two spectral lines indicate emission originating from distinct physical components. For our subsequent analysis, we did not extend our investigation to H$_2$CO, owing to its unresolved components and extensive prior studies \citep[e.g.][]{vanDishoeck1995, Ceccarelli2000, Loinard2000,vanderWiel2019}. Instead, our focus was directed towards the emission of carbon chains from the cold envelope of IRAS 16293.
The spectra of the detected c-C$_3$H and HC$_7$N lines are reported in Figures \ref{fig:spectra-C3H} and \ref{fig:spectra-HC7N}.
The spectral parameter of the observed lines, as well as the line integrated intensities used to calculate $N^{\rm tot}$ for each species, are reported in Tab \ref{Tab:lines}.
We calculate column densities for a temperature in the range 5--20 K following the results obtained by  \citet{Jaber2014} from the analysis of cyanopolyynes in the envelope of IRAS 16293.
The resulting $N^{\rm tot}$ are listed in Table~\ref{Tab:column-densities}. Note that, as for the L1544 case, the reported values are beam-averaged.

\subsubsection{The {\rm C$_n$X} family: {\rm c-C$_3$H}}

We detected four hyperfine components of the N=1$_{1,0}$--1$_{1,1}$ transition of c-C$_3$H (see Fig.~\ref{fig:spectra-C3H}), with upper level energy (E$_{\rm up}$)  of 0.71 K. 
From the LTE analysis of the velocity-integrated emission of the two available brightest lines, we derived $N^{\rm tot}_{\rm c-C_3H}$ = 1.0–1.4 $\times$  10$^{12}$ cm$^{-2}$, and 3.5–6.5 $\times$ 10$^{12}$ cm$^{-2}$, assuming a gas temperature between 5 and 20 K, respectively. 

The line profile associated with the brightest c-C$_3$H line (J = 3/2–3/2 F = 2–2) exhibits a red-shifted wing (up to $\sim$ 5 km s$^{-1}$). This is due to the fact that the GBT beam in X band (54$\arcsec$) encompasses the bright and extended molecular outflow emerging from the protostellar system.
Figure \ref{fig:IRAS-outflows} shows
the position of the GBT beam with respect the CO(3--2) outflow obtained using the Atacama Pathfinder Experiment (APEX, \citealt{2006gusten})\footnote[7]{https://www.apex-telescope.org} telescope by \citet{Kahle2023}. 
The present GBT observations sample a considerable part of the bright red-shifted lobe. This explains the red-shifted wing in the c-C$_3$H line (J = 3/2 to 3/2 F = 2-2) spectrum, which is also consistent with the CO(3--2)  spectrum extracted by \citet{Kahle2023} from the position of the red-shifted outflow peak.

We disentangled the contributions to the c-C$_3$H emission from the envelope and the outflow spectral wing by fitting the envelope with a gaussian profile and associating to the outflow the residual emission. We then derived their respective column densities. 
For the envelope, the integrated intensity is estimated using a Gaussian fit with FWHM of 0.5 km s$^{-1}$ centered on the systemic velocity of +4 km s$^{-1}$ (see Fig.~\ref{fig:spectra-C3H}). 
We obtained $N_{\rm tot}$ = 1--4 $\times$ 10$^{12}$ cm$^{-2}$, assuming a gas temperature between 5 and 20 K. 
The residuals are used to estimate the outflow column density which is in the range 2–4 $\times$ 10$^{12}$ cm$^{-2}$, adopting a temperature range of 20–40 K \citep{Jaber2017, Kahle2023}.
For the outflow component, using the H$_2$ column density derived from the C$^{17}$O column density of 1.9 $\times$ 10$^{15}$ cm$^{-2}$ measured by \citet{Kahle2023},
assuming an isotopic ratio of $^{16}$O/$^{17}$O = 2798 \citep{McKeegan2011}, and a CO/H$_2$ abundance ratio of 2 $\times$ 10$^{-4}$, we obtained an estimate of the c-C$_3$H abundance with respect to H$_2$ of about 10$^{-10}$.

\subsubsection{The polyynyl radicals family: {\rm C$_6$H}}

Despite the presence of transitions in the Ku band, no C$_6$H lines were detected in the observed frequency range.
Assuming a gas temperature between 5 and 20 K, we obtain an upper limit on the column density of $N_{\rm C_6H} \leq 1-5 \times 10^{11}$ cm$^{-2}$ from the line at 15248.2470 MHz.

\subsubsection{The cyanopolyynes family: {\rm HC$_7$N} and {\rm HC$_9$N}}

Figure~\ref{fig:spectra-HC7N} shows that, once smoothed the 
spectral resolution to $\sim$ 0.5 km s$^{-1}$, we report, for the first time,
the detection of HC$_7$N in IRAS 16293. 
The J = 13–12 transition was detected at 14.663 GHz with a S/N of 4$\sigma$ (see
Tab. \ref{Tab:lines}). From an LTE analysis of the line, we derive a column density in the range 6–11 $\times$ 10$^{10}$ cm$^{-2}$, assuming a gas temperature between 5 and 20 K. We also derived an upper limit on the HC$_9$N column density of $N_{\rm HC_9N} \leq 2.9-3.2 \times 10^{10}$ cm$^{-2}$ from the line at 15106.8910 MHz.



\begin{table*}
\begin{center}
\caption{List of the transitions observed at frequencies ($\nu_{\rm obs}$) different from those listed in the spectral catalogues ($\nu_{\rm spec}$), listed in Table 1. The uncertainties on the observed frequencies are assumed to be twice the spectral resolution of the GBT backend, i.e. 2.8 kHz. \label{Tab:freq}}
 \begin{tabular}{lc|cc|cc}
 \hline
 \hline
{} & {} & \multicolumn{2}{|c}{L1544} & \multicolumn{2}{|c}{IRAS 16293} \\
{Species} & {Transition} & \multicolumn{1}{|c}{$\nu_{\rm obs}$$^{\rm a}$} & {$\nu_{\rm obs}-\nu_{\rm spec}$} & \multicolumn{1}{|c}{$\nu_{\rm obs}$} & {$\nu_{\rm obs}-\nu_{\rm spec}$} \\  
{} & {} & \multicolumn{1}{|c}{(MHz)} & {(kHz)} & \multicolumn{1}{|c}{(MHz)} & {(kHz)} \\
\hline
c-C$_{\rm 3}$H & N = 1$_{\rm 1,0}$--1$_{\rm 1,1}$ J=1/2--1/2 F = 1-1 & 14686.634 &  +4   & - &  - \\
c-C$_{\rm 3}$H & N = 1$_{\rm 1,0}$--1$_{\rm 1,1}$ J=1/2--1/2 F = 0-1 & 14689.710 &  --8  & - &  - \\
c-C$_{\rm 3}$H & N = 1$_{\rm 1,0}$--1$_{\rm 1,1}$ J=1/2--3/2 F = 0-1 & 14755.378 &  --18 & - &  - \\
c-C$_{\rm 3}$H & N = 1$_{\rm 1,0}$--1$_{\rm 1,1}$ J=1/2--3/2 F = 1-2 & 14767.668 & --32  & 14767.670 & --30\\
c-C$_{\rm 3}$H & N = 1$_{\rm 1,0}$--1$_{\rm 1,1}$ J=1/2--3/2 F = 2-1 & 14812.014 &  +4   & - &  - \\
c-C$_{\rm 3}$H & N = 1$_{\rm 1,0}$--1$_{\rm 1,1}$ J=1/2--3/2 F = 1-1 & 14829.580 &  +9   & - &  - \\
c-C$_{\rm 3}$H & N = 1$_{\rm 1,0}$--1$_{\rm 1,1}$ J=3/2--3/2 F = 1-1 & 14895.247 &  +4   & 14895.253 &  +10 \\
c-C$_{\rm 3}$H & N = 1$_{\rm 1,0}$--1$_{\rm 1,1}$ J=3/2--3/2 F = 1-2 & 14910.615 &  --10 & - &  - \\
\hline
C$_{\rm 6}$H & J=7/2--5/2 $\Omega$=3/2 F=4--3 l=e & 9703.503 & --5 & - &  - \\
C$_{\rm 6}$H & J=7/2--5/2 $\Omega$=3/2 F=4--3 l=f & 9703.840 & +5  & - &  - \\
C$_{\rm 6}$H & J=11/2--9/2 $\Omega$=3/2 F=6--5 l=e & 15248.243 & --4 & - &  - \\
C$_{\rm 6}$H & J=11/2--9/2 $\Omega$=3/2 F=5--4 l=e & 15248.321 & --11 & - &  - \\
C$_{\rm 6}$H & J=11/2--9/2 $\Omega$=3/2 F=6--5 l=f & 15249.087 & +3  & - &  - \\
C$_{\rm 6}$H & J=11/2--9/2 $\Omega$=3/2 F=5--4 l=f & 15249.161 & +3 & - &  - \\
\hline
 \end{tabular}
 \end{center}
$^{\rm a}$ The observed frequencies were derived from the spectra at the native spectral resolution of 1.4 kHz.
\end{table*}

\subsection{Suggested corrections for selected c-C$_3$H, and C$_6$H rest frequencies}

In both L1544 and IRAS 16293 spectra, we detected several transitions of c-C$_3$H and C$_6$H at frequencies significantly offset from those reported in the spectroscopic databases, specifically, JPL for c-C$_3$H and CDMS for C$_6$H.
Thanks to the high spectral resolution of the present observations (1.4 kHz), we were able to accurately determine the discrepancies between the observed frequencies ($\nu_{\rm obs}$) with the laboratory values reported in the spectral catalogues ($\nu_{\rm spec}$).
Table~\ref{Tab:freq} reports the observed frequencies for the c-C$_3$H, and C$_6$H transitions where the discrepancy is larger than twice the spectral resolution. Note that, when observed in both L1544 and IRAS 16293,
the frequency shifts are consistent,
supporting the reliability of the corrections.
We then propose these corrections in future updates of these databases.
The corrected values for c-C$_3$H are in agreement with the recent laboratory measurements reported by \citet{2025Xue}.
For C$_6$H, we examined the new catalog reported by \citet{Remijan2023}, and the transitions considered show the same frequency shifts as those found in our observations.


\section{Discussion}
\label{sec:discussion}


\subsection{Comparison with previous observations}
\label{sec:discussion-prev-obs}

\textbf{The L1544 prestellar core:} L1544 has long been recognized as a chemically rich environment, particularly abundant in small carbon chains (C$_2$S, C$_3$S, C$_3$N, and c-C$_3$H), and cyanopolyynes \citep[e.g.][]{Lin2022,Giers2022,Bianchi2023}. 
The present GBT observations further enrich this picture, revealing carbon chains such as C$_2$S, C$_3$S, C$_3$N, and c-C$_3$H, along with the polyynyl radicals C$_4$H and C$_6$H, which are known to be key precursors of cyanopolyynes \citep{loison2014,Giani2025}.
C$_2$S was first observed by \citet{Suzuki1992}, with more recent estimates of the column density derived by \citet{2018vastel-Sbearing} and \citet{nagy2019} using the IRAM 30-m telescope. Our derived column densities are systematically a factor of 2 higher, but remain in agreement with these studies within the associated uncertainty.
This may suggest that the IRAM 30-m observations with telescope HPBWs between 24$\arcsec$ and 30$\arcsec$, centered on the dusty peak of L1544, probably miss the extended emission which is instead recovered by the GBT beam. Indeed, our line profiles are consistent with the C$_2$S emission map reported by \citet{Spezzano2017}, which shows an asymmetric distribution with enhanced emission in the south-east region of L1544.
C$_3$S and C$_3$N detections were previously reported by \citet{2018vastel-Sbearing}, and \citet{Vastel2019} using the IRAM 30-m antenna, finding column densities compatible (within a factor of 2) with the present ones.

The cyclic isomer c-C$_3$H was first detected in L1544 by \citet{Mangum1990} using the NRAO 43m telescope.
In our dataset, only the cyclic form (c-C$_3$H) is detected and not the linear one (l-C$_3$H).
\citet{Loison2017} analyzed the chemistry of c- and l-C$_3$H in dark clouds (TMC-1 and B1-b), reporting typical c/l abundance ratios of $\sim$5. Thanks to calculations of rovibrational density of states, they showed that, when formed in highly excited states, the isomerization between the cyclic and linear isomers (i.e. c-C$_3$H $\rightleftarrows$ l-C$_3$H) favor the formation of the cyclic form.
In addition, they found that the dissociative recombination (DR) of c,l-C$_3$H$_2^+$ leads mainly to cyclic c-C$_3$H.
Although the formation of c-C$_3$H is generally favored, the chemistry of the two isomers is tightly coupled. Given their common precursors and the ease of interconversion, c- and l-C$_3$H likely coexist and trace similar physical regions, and both should be considered in astrochemical modeling. As a matter of fact, our observed line profiles for c-C$_3$H are consistent with the l-C$_3$H map obtained by \citet{Spezzano2017} with the IRAM 30-m, revealing a more uniform distribution than C$_4$H and C$_2$S.  
The detection and analysis of cyanopolyynes, up to HC$_9$N, in L1544 was previously reported by \citet{Bianchi2023}, who derived abundance ratios of HC$_5$N:HC$_7$N:HC$_9$N = 1:0.16:0.25. 
C$_4$H and C$_6$H were previously observed in L1544 by \citet{Gupta2009} using GBT in the 18–22 GHz frequency range.
Although our column density estimate for C$_6$H is consistent with their results, our value for C$_4$H is almost one order of magnitude lower, leading to a significantly higher C$_6$H/C$_4$H ratio (4–6$\%$ compared to 0.5–0.9$\%$). This discrepancy arises because our analysis adopts the revised dipole moment for C$_4$H from \citet{oyama2020}, which is 2.4 times higher than the earlier value (2.1 D vs. 0.9 D) listed in the JPL and CDMS spectral databases. This leads to C$_4$H abundances that are a factor of 5.44 lower than previous estimates, resulting in proportionally higher C$_6$H/C$_4$H ratios.

\textbf{The IRAS 16293 protostellar system:} IRAS 16293 has been a subject of extensive investigations, with studies focusing on both its cold envelope \citep[e.g.][]{Blake1994,vanDishoeck1995, Caux2011} and the protostellar system \citep[e.g.][]{Jorgensen2016,Jorgensen2018}.
The narrow linewidths (FWHM $\leq$ 1 km s$^{-1}$) and low upper level energies of the molecular lines observed with this study, as well as the analysis of H$_2$CO lines, confirm that the detected emission predominantly arises from cold envelope gas, with minor contamination from the red-shifted outflow (e.g., in c-C$_3$H; see Sec. \ref{subsec:IRAS-LTE} and Fig. \ref{fig:spectra-C3H}).
Our observations allow for the first detection of HC$_7$N towards IRAS 16293. We also report, for the first time, upper limits for C$_6$H and HC$_9$N. 
Previous detections of c-C$_3$H and C$_4$H were reported by \citet{Caux2011}. The emission is confirmed to originate from the cold envelope due to its narrow line profiles; however, no column density was derived. Emission from C$_4$H have been reported also by \citet{Sakai2009} and \citet{Lindberg2016}, using the Mopra 22-m and the Kitt Peak 12-m telescopes, respectively. Based on the C$_4$H column density from \citet{Lindberg2016} and the C$_6$H column density derived in this work, we derive a C$_6$H/C$_4$H ratio of approximately 0.05, consistent with values observed in other sources (see Sec.~\ref{subsec:discussion-IRAS} and \ref{subsec:model}).

Beside the upper limit on CCS column density, $N^{\rm tot}$$\leq$1$\times$10$^{15}$ cm$^{-2}$, set by \citet{dishoeck1995} using the CSO 1.4-m and JCMT 15-m antennas, and $N^{\rm tot}$$\leq$4$\times$10$^{13}$ cm$^{-2}$ reported by \citet{2018Drozdovskaya} with ALMA in the IRAS 16293 B hot corino,
no additional observational data are available for CCS, C$_3$S, or C$_3$N in the IRAS 16293 envelope for a direct comparison with the present dataset.

HC$_3$N has previously been detected (i) toward protostars using SMA, VLA, and ALMA interferomets by \citet{Kuan2004}, \citet{Chandler2005}, and \citet{calcutt2018}, respectively, and (ii) sampling large spatial scales  by \citet{dishoeck1995} using JCMT 15m (beam size 20$\arcsec$) and \citet{Jaber2017} using IRAM 30-m (beam size 30$\arcsec$). 
\citet{dishoeck1995} observed three high-excitation transitions (28–27, 27–26, and 26–25), deriving a rotational temperature of $\sim$115~K and a column density of $N^{\rm tot}_{\rm HC_3N}$ = 2--5 $\times$ 10$^{12}$ cm$^{-2}$. In contrast, \citet{Jaber2017} observed 15 transitions in addition to the three lines reported by \citet{dishoeck1995}, covering also transitions from (9-8) up to (30–29). 
Based on these data, they derived an excitation temperature of 20~K and a column density of $N^{\rm tot}_{\rm HC_3N}$ = 3.2--5.6 $\times$ 10$^{11}$ cm$^{-2}$ for the outer envelope.
Since the high-excitation lines detected by \citet{dishoeck1995} are not expected to arise from the cold envelope, we adopt the values reported by \citet{Jaber2017} for the outer envelope in the following analysis. In addition, \citet{Jaber2017} also reported the detection of HC$_5$N in the envelope of IRAS 16293, deriving a column density of $N^{\rm tot}_{\rm HC_5N}$ = 3.6--7 $\times$ 10$^{11}$ cm$^{-2}$ at 20~K.
The HC$_7$N column density derived in this work is consistent with the upper limit (N$\leq$5.3$\times$10$^{13}$ cm$^{-2}$) derived by \citet{Jaber2017}.
This implies no significant beam dilution differences are observed between GBT and IRAM 30-m data, implying an extended emitting region, larger than both telescope beams.
The inferred HC$_5$N:HC$_7$N:HC$_9$N ratios in the IRAS 16293 envelope is 1:0.14:<0.28, closely matching those derived for L1544 \citep{Bianchi2023}, suggesting similar cyanopolyyne chemistry across different cold environments.

\begin{figure}
\includegraphics[scale=0.58]{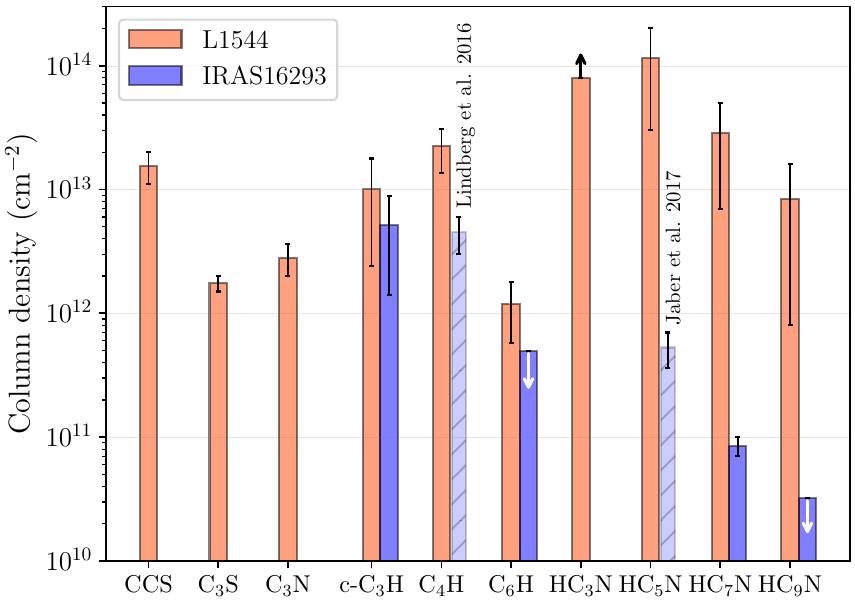}
  \caption{Column densities of  carbon chains and cyanopolyynes (see Tab. \ref{Tab:column-densities}) observed in L1544 (orange bars) and IRAS 16293 (blue bars). The upper limits on C$_6$H and HC$_9$N column densities in L1544 are shown with white arrows.
   The black arrow is for the lower limit on the HC$_3$N column density due to opacity effects \citep{Bianchi2023}.
 The C$_4$H and HC$_5$N values in IRAS 16293 are taken from \citet{Lindberg2016} and \citet{Jaber2017}, respectively.
  } 
  \label{fig:histogram}
\end{figure}


\subsection{Comparison between prestellar core and protostellar envelope chemistry}\label{subsec:discussion-IRAS}

\begin{figure}
\includegraphics[scale=0.51]{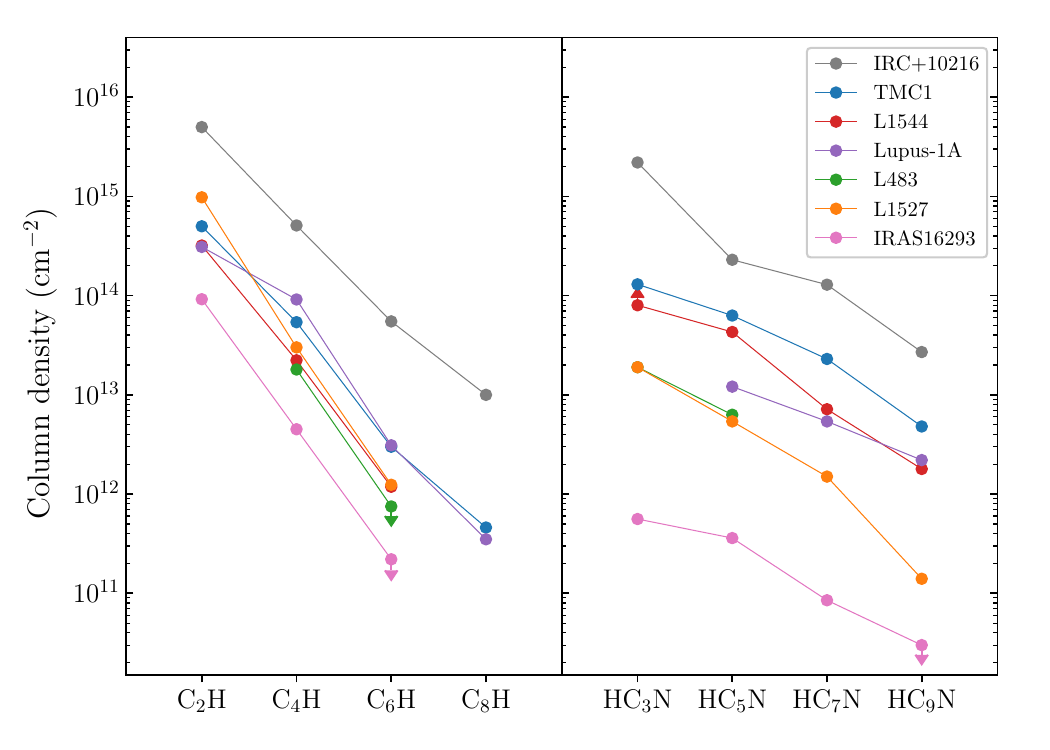}
  \caption{Comparison between column densities of polyynyl radicals C$_{\rm 2n}$H (Left panel) and cyanopolyynes HC$_{\rm 2n+1}$N (Right panel) measured in different star-forming regions. Values for L1544 are from the present work, and from \citet{Bianchi2023,giers2023}, while values for IRAS 16293 are from the present work, and from \citet{Jaber2014,Lindberg2016}.
  Values for sources different than L1544, and IRAS 16293  are taken from \citet[][and references therein]{oyama2020}.
  } 
  \label{fig:trend-coldens}
\end{figure}

To investigate the chemical differentiation between L1544 and IRAS 16293, we compared the column densities of carbon chains and cyanopolyynes in the two sources (Fig. \ref{fig:histogram}). The comparison reveals several key points:
\begin{itemize}
\vspace{-5pt}
    \item[(i)] Cyanopolyynes exhibit similar relative column density ratios in both sources (HC$_5$N:HC$_7$N:HC$_9$N = 1:0.16:0.25 in L1544 and 1:0.14:<0.28 in IRAS 16293) yet their absolute column densities are systematically lower in IRAS 16293, possibly related to a chemical difference between the cold molecular envelope of IRAS 16293 and L1544;
    \item[(ii)] A similar trend is observed for the polyynyl radicals (e.g., C$_4$H, C$_6$H), consistent with their role as precursors of cyanopolyynes;
    \item[(iii)] c-C$_3$H is the only species with comparable column densities in both sources, within uncertainties;
    \item[(iv)] No detections of C$_2$S, C$_3$S, or C$_3$N are available in the literature for the IRAS 16293 envelope, preventing a direct comparison for these species.
\end{itemize}
In Figure~\ref{fig:trend-coldens} we compare the column densities of polyynyl radicals C$_{\rm 2n}$H and cyanopolyynes HC$_{\rm 2n+1}$N measured in different star forming regions. 
The figure shows that, using the updated C$_4$H spectral values, a clear exponential decrease in the column densities of polyynyl radicals is revealed. A similar, though shallower, trend is observed for cyanopolyynes, supporting a chemical connection between the C$_{\rm 2n}$H and HC$_{\rm 2n+1}$N families. 
Unsaturated carbon chains are believed to form mainly through neutral-neutral and ion-neutral reactions in the gas phase. Formation on grain surfaces is unlikely due to rapid hydrogenation by accreting H atoms \citep{hiraoka2000,kobayashi2017,molpeceres2022,fedoseev2025,raaphorst2025}. \citet{Giani2025} revised the neutral-neutral reactions of formation of HC$_5$N and found that in cold environments (T$\sim$10 K) reactions such as C$_6$H + N, C$_3$N + C$_2$H$_2$, and HC$_3$N + C$_2$H are the dominant ones.
The observed exponential decrease in the abundances of polyynyl radicals and cyanopolyynes likely reflects a stepwise chemical growth, in which larger species derive from smaller precursors. For example, HC$_5$N may originate from HC$_3$N and C$_6$H, which themselves derive from smaller species such as C$_4$H. 
In general, the formation of carbon chains requires large amounts of atomic carbon. So far, two main scenarios have been proposed: either the gas is chemically young and carbon is not yet locked into CO, or a large irradiation by UV photons or cosmic rays drives CO dissociation, releasing carbon atoms into the gas phase \citep[e.g.,][]{agundez2013,Sakai2013,Spezzano2016b,Spezzano2017}.
Large values of $\zeta_{\rm CR}$(>10$^{-16}$ s$^{-1}$) are however excluded by observations in L1544, as models better reproduces ions abundances when adopting $\zeta_{\rm CR}$=2-3$\times$10$^{-17}$ s$^{-1}$ \citep{2021Redaelli}.
To further explore the origin of the chemical differences between the two sources, we performed astrochemical modeling with varied physical parameters. The model description and results are presented in the following section.

\begin{table}
    \centering
        \caption{Physical parameters and initial elemental abundances used in the cold prestellar core/envelope model.
    The upper half table lists the adopted parameters: H density, n$_{\rm H}$, temperature, T, cosmic-ray ionisation rate, $\zeta_{\rm CR}$, visual extinction, A$_{\rm v}$, dust grain radius, a$_{\rm d}$, and
    grain density, $\rho_{\rm d}$.
    The lower half table lists the initial elemental abundances relative to H nuclei, adapted from \citet{jenkins2009unified}.
    }
    \begin{tabular}{cc|cc}
        \hline
        \hline
        \multicolumn{4}{c}{Physical parameters of the cold gas} \\
        Parameter & Value & Parameter & Value \\
        \hline
        n$_{\rm H}$    [cm$^{-3}$]   & $2\times 10^4$    &   A$_{\rm v}$ [mag]  & 1-10 \\
        T            [K]         & 10                 &  a$_{\rm d}$        [$\mu$m]    & 0.1   \\
        $\zeta_{\rm CR}$ [s$^{-1}$]  & $0.1-10\times 10^{-16}$ & $\rho_{\rm d}$          [g~cm$^{-3}$]  & 3  \\
        \hline
        \hline
        \multicolumn{4}{c}{Initial elemental abundances (wrt H)} \\
        Element & Abundance & Element & Abundance \\
        \hline
H$_2$   & $5.0 \times 10^{-1}$  & Si$^+$  & $8.0 \times 10^{-9}$  \\  
He      & $9.0 \times 10^{-2}$  &  P$^+$   & $2.0 \times 10^{-10}$ \\
C$^+$   & $2.0 \times 10^{-5}$  &   Na$^+$  & $2.0 \times 10^{-9}$  \\
O       & $2.6 \times 10^{-5}$  &   Fe$^+$  & $3.0 \times 10^{-9}$  \\
N       & $6.2 \times 10^{-6}$  &   Cl$^+$  & $1.0 \times 10^{-9}$  \\
S$^+$   & $8.0 \times 10^{-8}$  &   F$^+$   & $1.0 \times 10^{-9}$  \\
        \hline
    \end{tabular}
    \label{tab:model-abd-param}
\end{table}

\subsection{Astrochemical modeling}\label{subsec:model}

\begin{figure*}
\includegraphics[scale=0.58]{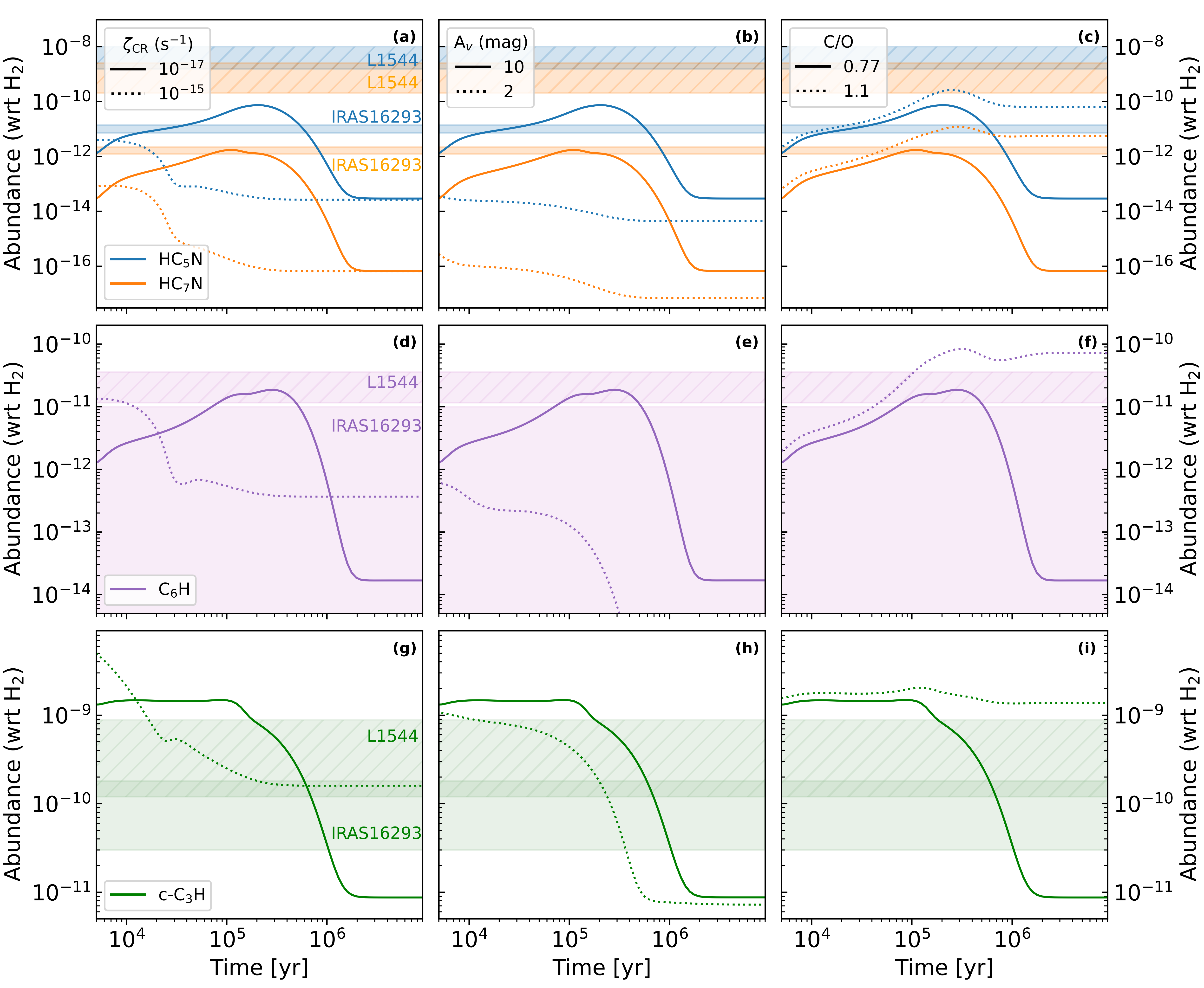}
\caption{Comparison between the observed and modeled abundances of HC$_5$N, HC$_7$N, C$_6$H and c-C$_3$H for different $\zeta_{\rm CR}$, A$_{\rm v}$, and C/O ratios. Panels (a, b, c) show the predicted abundances of HC$_5$N (blue lines) and HC$_7$N (orange lines) as a function of the parameters indicated in the legends. Panels (d, e, f) and (g, h, i) show the same for C$_6$H (purple lines) and c-C$_3$H (green lines), respectively.
Observed values from L1544 and IRAS 16293 are shown as shaded bands, using the same color code as the model lines; dashed hatching indicates L1544.
We assumed an N(H$_2$) of 2$\times$10$^{22}$ cm$^{-2}$ for L1544 and 5$\times$10$^{22}$ cm$^{-2}$ for IRAS 16293 (based on Herschel observations, see text).} 
  \label{fig:fig-CR-Av-CO-5-7}
\end{figure*}

\subsubsection{Model description}

We ran the MyNahoon astrochemical model \citep{wakelam2005estimation,wakelam2010sensitivity} to investigate the chemical diversity between L1544 and the cold molecular envelope of IRAS 16293.  The model has been extensively described in previous works \citep{Podio2014,codella2017,giani2023revised}. In summary, it calculates the gas phase abundances of $\sim$ 500 molecules as a function of time for some specific physical parameters, such as gas temperature, T, volume density, n$_{\rm H}$, cosmic-ray ionization rate, $\zeta_{\rm CR}$, and visual extinction, A$_{\rm v}$. 
We adopted the GRETOBAPE gas-phase reaction network \citep{tinacci2023-gretobape}, including recent updates on the chemistry of formation of HC$_5$N \citep{Giani2025} and c-C$_3$H \citep{Loison2017}. Surface chemistry (including adsorption and desorption) is not treated in the network, except for the formation of H$_2$ in dust grains. As unsaturated carbon chains are formed primarily in the gas phase, we do not expect a significant impact from the missing grain-surface processes. The initial elemental abundances and physical parameters adopted in the simulations are summarized in Tab.~\ref{tab:model-abd-param}. 
To simulate the evolution of a typical translucent cloud into a molecular cloud, we assumed as initial conditions that elements with ionization potentials below 13.6 eV (e.g., C, S, Si, P) are ionized, while those with higher values (N, O) are neutral.
In all simulations, hydrogen is assumed to be fully molecular.
The 0D model cannot capture the asymmetric spatial distributions of different species, likely arising from the source's physical structure \citep{Jensen2023}. However, we account for this limitation by comparing beam-averaged abundances with those predicted by the model. The key novelty of this model lies in the updated chemical network with thoroughly studied reaction rates, making it a valuable tool for exploring how the abundances of long carbon chains vary with different parameters. 
To assess the influence of cosmic rays and UV radiation on carbon chain formation, we ran a grid of models assuming the following ranges:
(i) $\zeta_{\rm CR}$ between $1\times 10^{-17}$~s$^{-1}$ and $1\times 10^{-15}$~s$^{-1}$ \citep{2014ceccarellib,2021Redaelli,2025redaelli,2023sabatini,2024socci};
(ii) A$_{\rm v}$ between 1 and 10 mag.
(iii) C/O elemental abundance ratios between 0.77 and 1.1.
Since elemental depletion in the outer envelope of IRAS 16293 may not yet be complete, we also ran a model in which the initial elemental abundances from Tab.~\ref{tab:model-abd-param} were increased by a factor of 10.
In order to compare the observed column densities with the abundances predicted by the model, we consider the H$_2$ column density to be at least 2 $\times$ 10$^{22}$ cm$^{-2}$ and 5 $\times$ 10$^{22}$ cm$^{-2}$ for L1544 and IRAS 16293, respectively, from Herschel observations \citep{Kirk2013, Spezzano2016b, Ladjelate2020} and in agreement with previous estimates \citep{Crapsi2005, vanDishoeck1995}. 
We note that these values correspond to the integrated H$_2$ column density along the line of sight, while the emission of carbon chains and cyanopolyynes may originate from an external layer, where the H$_2$ column density is likely lower. No direct measurements of H$_2$ in these region are available, making it difficult to accurately determine molecular abundances. To avoid assuming an arbitrarily reduced H$_2$ column density, we adopted the integrated values while emphasizing that the resulting abundances reported in Tab.~\ref{Tab:column-densities} should be used with caution. Accordingly, in the discussion of the chemical modeling, we focus primarily on abundance ratios rather than absolute abundances.


\subsubsection{Abundance of carbon chains}

As a first test, we examined how variations in A$_{\rm v}$, $\zeta_{\rm CR}$, and C/O elemental ratio affect the predicted abundances of carbon chains. For these models, we adopted the same gas density and temperature for L1544 and IRAS16293, given their similar physical conditions. Figure~\ref{fig:fig-CR-Av-CO-5-7} compares the observed and modeled abundances of HC$_5$N, HC$_7$N, C$_6$H, and c-C$_3$H under these varying conditions.
For both sources the best agreement with observations is obtained for cosmic-ray ionization rate $\zeta_{\rm CR}=1\times 10^{-17}$~s$^{-1}$, visual extinction A$_{\rm v}$=10 mag and C/O=0.77. 
In all models, cyanopolyynes are underproduced by approximately two orders of magnitude compared to L1544 values. Conversely, good agreement is achieved for C$_6$H and c-C$_3$H abundances at times around 2-3 $\times$ 10$^5$ years. The recent revision of the gas-phase formation reactions of HC$_5$N by \citet{Giani2025} suggested that the C$_6$H + N reaction is the most important in cold regions. 
However, while the model considering this updated network correctly reproduces the C$_6$H abundance, it fails to reproduce that of HC$_5$N. If C$_6$H were indeed the major HC$_5$N precursor, the model should accurately predict HC$_5$N abundances. This discrepancy strongly suggests that additional formation reactions are missing from the chemical network, not only for HC$_5$N but also for the longer cyanopolyynes.
In IRAS 16293, the observed abundances of c-C$_3$H and HC$_5$N are well reproduced by the model at an age of approximately 7$\times$10$^5$ years, in line with the source being more chemically evolved than L1544.
At this time, however, the model underestimates the abundance of HC$_7$N by more than one order of magnitude.
Unfortunately, the upper limit on the C$_6$H abundance is not sufficiently constraining for a meaningful comparison.
When adopting less depleted initial elemental abundances, all species show comparable abundances at $10^5$ years (with variations of less than one order of magnitude). At later times (t>10$^6$ years), however, their abundances decrease by several orders of magnitude relative to the predictions obtained with the more depleted abundances, thereby worsening the agreement with the observations.

\begin{figure}
\includegraphics[scale=0.54]{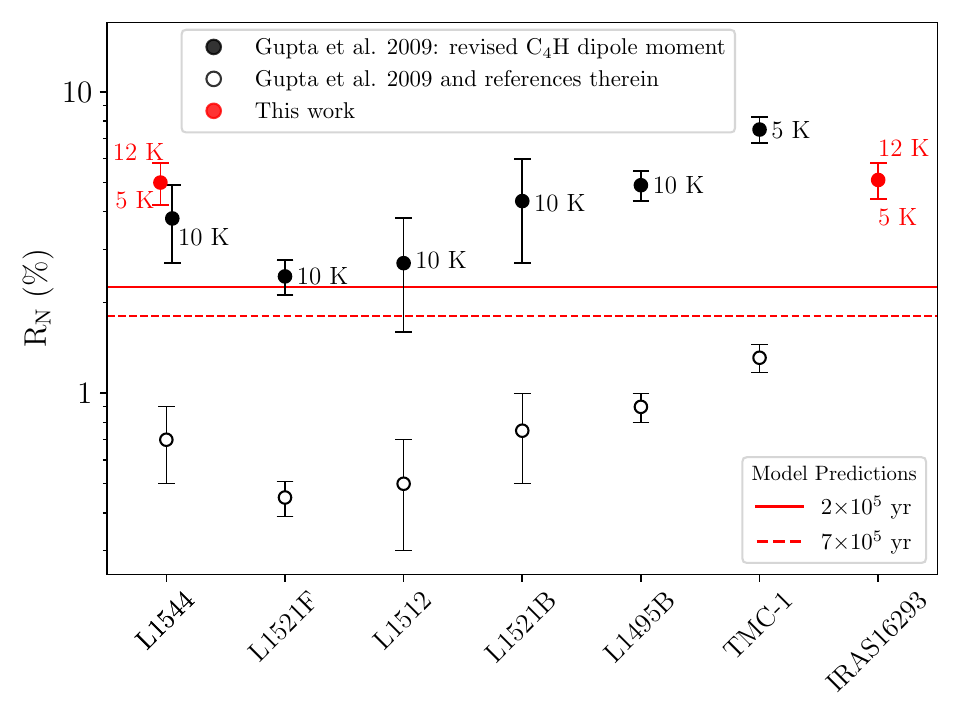}
\caption{Comparison of the C$_6$H/C$_4$H ratio (R$_{\rm N}$) obtained for L1544 with those derived in cold molecular clouds and protostars. Open circles correspond to data from \citet[][and references therein]{Gupta2009}, while filled circles correspond to the same data corrected taking into account the revised dipole moment of C$_4$H (see text in Sec.~\ref{sec:discussion} and \citealt{oyama2020}). The R$_{\rm N}$ value derived in this work for L1544 and IRAS 16293 are shown in red. The temperatures at which the column densities of C$_4$H and C$_6$H have been derived are reported next to each point. R$_{\rm N}$ values predicted by the model at times of 2$\times$10$^5$ and 7$\times$10$^5$ years are shown as solid and dashed red lines, respectively.
Model predictions are obtained assuming $\zeta_{\rm CR}$=$10^{-17}$~s$^{-1}$ and A$_{\rm v}$=10 mag.
} 
  \label{fig:C4H/C6H-model+obs}
\end{figure}

\subsubsection{Abundance ratios of carbon chains}

To overcome the substantial uncertainties associated with observed H$_2$ column densities, we also compare the model predictions with observed abundance ratios of the different species.
Figure~\ref{fig:C4H/C6H-model+obs} shows the comparison of the predicted C$_6$H/C$_4$H ratios with those derived from observations, using both the old and revised dipole moment of C$_4$H. As discussed in Sec.~\ref{subsec:discussion-IRAS}, \citet{oyama2020} estimated the C$_4$H dipole moment to be 2.4 times larger than the previous value, implying that previous estimates of its column density were overestimated by a factor of $\sim$6. 
When the corrected dipole moment is used, the observed C$_6$H/C$_4$H ratios, including those reported in this work, are in agreement with model predictions within a factor 3. 
This possible discrepancy highlights the need for a revision of the formation and destruction pathways for these species.

\begin{figure}
\includegraphics[scale=0.55]{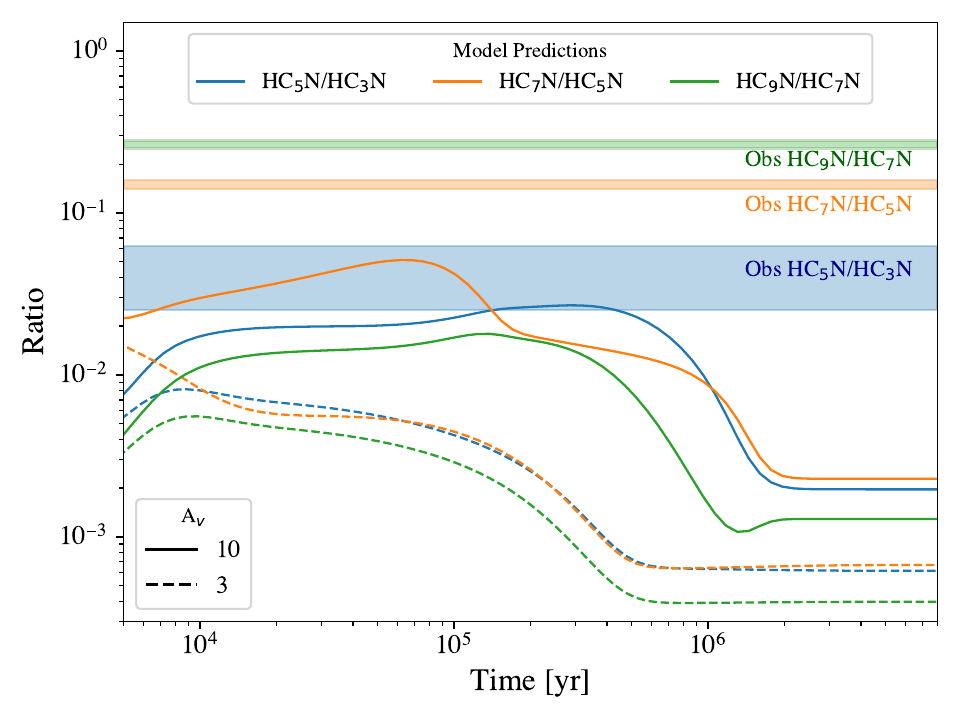}
\caption{Comparison between predicted and observed abundance ratios of HC$_5$N/HC$_3$N, HC$_7$N/HC$_5$N, and HC$_9$N/HC$_7$N for different visual extinctions (A$_{\rm v}$). Model predictions are shown as lines: blue for HC$_5$N/HC$_3$N, orange for HC$_7$N/HC$_5$N, and green for HC$_9$N/HC$_7$N. Solid lines correspond to A$_{\rm v}$=10, while dashed lines to A$_{\rm v}$=3. The observations, based on values derived for both L1544 and IRAS 16293 in this work, are represented as shaded bands in the same colors.
Model predictions are obtained assuming $\zeta_{\rm CR}$=$10^{-17}$~s$^{-1}$ and A$_{\rm v}$=10 mag.
} 
  \label{fig:ratio-579}
\end{figure}

\begin{figure}
\includegraphics[scale=0.49]{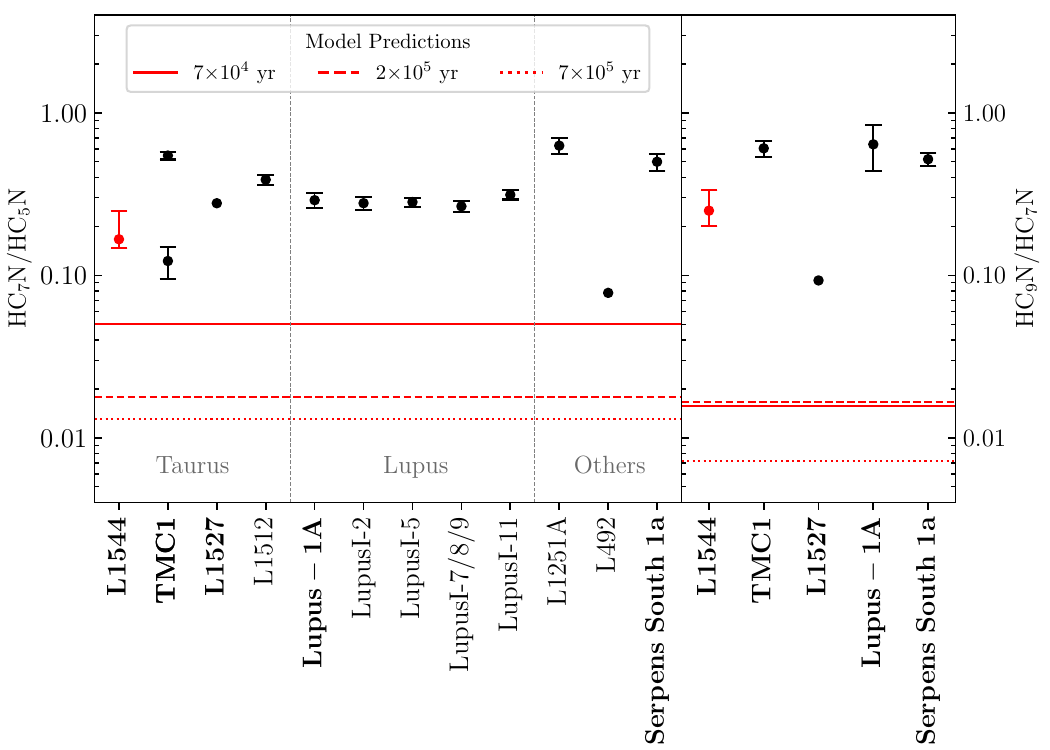}
\caption{Comparison of the HC$_7$N/HC$_5$N (Left panel), and HC$_9$N/HC$_7$N (Right panel) ratios obtained for L1544 with those derived in cold molecular clouds. The values derived in this work for L1544 are shown as red circles. 
Values for L1544 are from the present work, while values for sources different than L1544 are taken from \citet[][and references therein]{Bianchi2023}.
 Values predicted by the model at times of 7$\times$10$^4$, 2$\times$10$^5$ and 7$\times$10$^5$ years are shown as solid, dashed and dotted red lines, respectively.
 Model predictions are obtained assuming $\zeta_{\rm CR}$=$10^{-17}$~s$^{-1}$ and A$_{\rm v}$=10 mag.
 Sources for which both the HC$_7$N/HC$_5$N and HC$_9$N/HC$_7$N ratios are available are in boldface. 
 Labels indicate the regions where sources are located: Taurus, Lupus, or other regions (Chameleon, Cepheus, and Aquila Rift).
 } 
  \label{fig:comp-579-ratios}
\end{figure}

\begin{figure}
\includegraphics[scale=0.53]{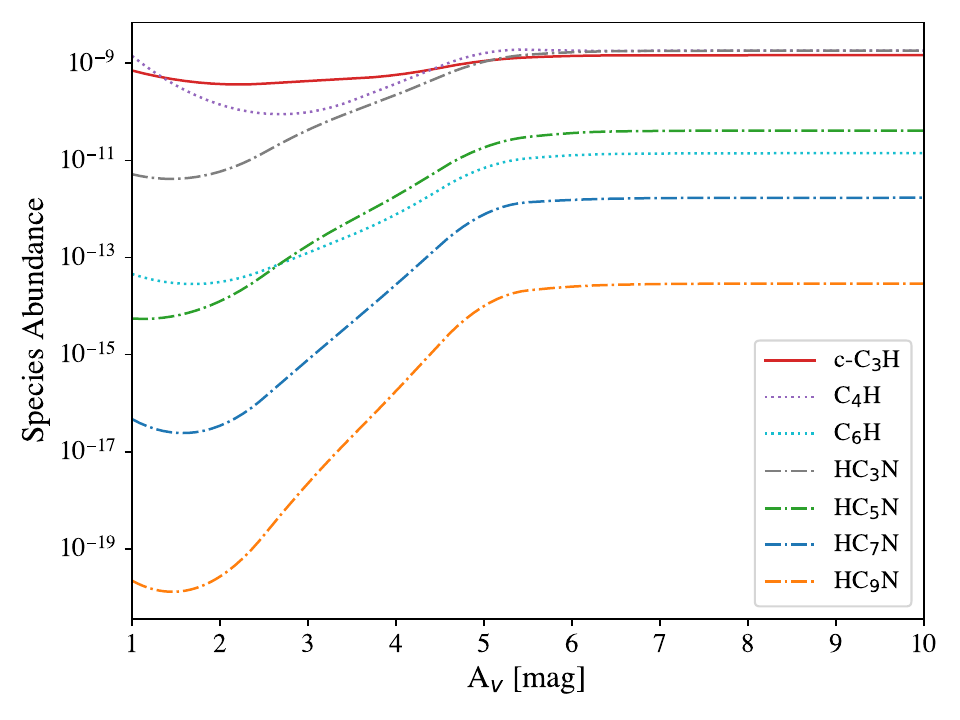}
\caption{Predicted abundances (with respect to H$_2$) of c-C$_3$H (red solid line), C$_4$H (purple dotted line), C$_6$H (cyan dotted line), HC$_3$N (grey dashed line), HC$_5$N (green dashed line), HC$_7$N (blue dashed line), and HC$_9$N (orange dashed line) as a function of A$_{\rm v}$. The model assumes n$_{\rm H_2}$=10$^4$ cm$^{-3}$, T=10~K and $\zeta_{\rm CR}$=10$^{-17}$ s$^{-1}$. The predictions refer to the model at 10$^5$ yr.
} 
  \label{fig:PDR-abd-all}
\end{figure}

Fig.~\ref{fig:ratio-579} compares the predicted HC$_5$N/HC$_3$N, HC$_7$N/HC$_5$N and HC$_9$N/HC$_7$N ratios with those derived from the observations presented in this work. 
There is a clear discrepancy between models and observations with the exception of the HC$_5$N/HC$_3$N ratio, which is reasonably reproduced at t$\sim$2-3$\times$10$^5$ years with A$_{\rm v}$=10 mag.
Figure~\ref{fig:comp-579-ratios} further illustrates that this discrepancy is not limited to a single source: even across different regions with different evolutionary ages, current models consistently underestimate the HC$_7$N/HC$_5$N and HC$_9$N/HC$_7$N ratios. Specifically, the predicted HC$_7$N/HC$_5$N ratios are lower than observed by factors of $\sim$2–10, while the HC$_9$N/HC$_7$N ratios are underestimated by factors of $\sim$7–50.
Although existing models do reproduce the general trend of decreasing abundance with increasing molecular size, they fail to reproduce the observed abundance ratios, particularly for the longest chains. 
%
%
In order to explore the possibility that cyanopolyynes and carbon chains arises from a UV illuminated region, as proposed by \citet{Spezzano2016b} for L1544, we verified the impact of different A$_{\rm v}$ values on the model predictions.
Fig.~\ref{fig:PDR-abd-all} shows the predicted abundances (with respect to H$_2$) of various carbon chains as a function of the visual extinction A$_{\rm v}$. We conservatively assume a gas volume density n$_{\rm H_2}$=10$^4$ cm$^{-3}$.
Using the current network, cyanopolyynes and polyynyl radicals show a strong decrease by 1 to 4 orders of magnitude from high to low A$_{\rm v}$. In contrast, the abundance of c-C$_3$H remains nearly constant across the entire A$_{\rm v}$ range (between 1 and 10).
Given that the major formation routes for cyanopolyynes in the current chemical network are neutral-neutral reactions, it is not surprising that our model predicts a decrease in cyanopolyyne abundance instead of an increase for higher UV illumination. On the contrary, the model successfully explains the presence of small carbon chains, such as c-C$_3$H and c-C$_3$H$_2$, in highly irradiated environments \citep{Favre2018}. This suggests that
additional formation reactions for cyanopolyynes and their precursors are missing in the network. More specifically, additional ion–molecule reactions could explain the enhancement of cyanopolyynes in UV exposed environments.

\subsubsection{Observations vs models: carbon chains formation routes}

Following the comparison of model predictions and observations reported in the previous sections, our main findings are:
\begin{itemize}
    \vspace{-5pt}
    \item[(i)] Models reproduce c-C$_3$H and C$_6$H in L1544 for Av=10, standard $\zeta_{\rm CR}$ and C/O values, and times of $\sim$ 2-3 $\times$ 10$^5$ yr. For the same conditions, cyanopolyynes are underproduced by a factor $\sim$100. 
    For IRAS 16293, c-C$_3$H and HC$_5$N are reproduced by the model for A$_{\rm v}$=10 and times of $\sim$ 7 $\times$ 10$^5$ years, but not HC$_7$N.
        \vspace{2pt}
    \item[(ii)] Models reasonably reproduce the abundance ratios between species within the same chemical family for A$_{\rm v}$=10 and evolutionary timescales in the range 2-7 $\times$ 10$^5$ years. The C$_6$H/C$_4$H and HC$_5$N/HC$_3$N ratios are reproduced in both L1544 and IRAS 16293 (see Fig.~\ref{fig:C4H/C6H-model+obs} and \ref{fig:ratio-579}), while the HC$_7$N/HC$_5$N and HC$_9$N/HC$_7$N ratios are underestimated by a factor of $\sim$2--50 likely due to the incomplete chemical network for large cyanopolyynes (see Fig.\ref{fig:comp-579-ratios});
        \vspace{2pt}
     \item[(iii)] Cyanopolyynes are predicted to be abundant in young objects, while their abundances decrease at later evolutionary stages (t>3$\times$10$^5$ years; see Fig.~\ref{fig:fig-CR-Av-CO-5-7}). As proposed for the starless core TMC-1, this behavior can be explained by the larger availability of carbon atoms before being converted into CO  \citep{agundez2013}. This explains the higher cyanopolyyne abundances observed in L1544 compared to IRAS 16293, since L1544 is a younger source.
        \vspace{2pt}
    \item[(iv)] Current models do not support the hypothesis that cyanopolyyne abundances are enhanced by elevated cosmic ray ionization rate or intense UV irradiation, despite indications from the asymmetric spatial distribution of carbon chains and cyanopolyynes observed towards the southeastern region of L1544 \citep{Spezzano2016b, Bianchi2023}.  The discrepancy between predicted and observed abundances is particularly pronounced for cyanopolyynes. This highlights the incompleteness of current chemical networks, especially regarding their formation reactions, which in turn impedes a comprehensive understanding of their primary formation mechanisms.
    \item[(v)] Although previous studies have shown that neutral–neutral reactions are the primary formation routes of cyanopolyynes \citep{Giani2025}, they cannot fully explain the observed abundances, especially in L1544. We therefore strongly suggest that additional ion–molecule reactions, beyond those currently included in the networks, are involved, as the production of ionic species can be enhanced in highly irradiated environments. A thorough revision of the formation pathways of the HC$_5$NH$^+$ ion, which could be a major precursor of HC$_5$N, is needed to verify this hypothesis (see Appendix for more details). 
\end{itemize}

\subsection{Future impact of SKA observations on carbon chain chemistry}

The present observations confirm that complex carbon chemistry is already active from the earliest stages of Sun-like planetary system formation, with these species potentially contributing to the synthesis of molecules of prebiotic relevance.
However, single-dish facilities have tipically an angular resolution of 54$\arcsec$-84$\arcsec$ in the 8-15 GHz range (corresponding to $\sim$10~000 au scales for nearby star-forming regions), which allow for the observation of these species in extended sources, such as prestellar cores or protostellar envelopes, leaving it unclear whether they also exist at smaller angular scales.
New possibilities to observe these species at high-angular resolution will be offered by next-generation facilities such as MeerKAT\footnote[8]{www.sarao.ac.za/science/meerkat/} and the Square Kilometre Array Observatory (SKAO)\footnote[9]{www.skao.int/en}. Specifically, MeerKAT Band 5b (covering 8.3–15.4 GHz), currently under development, will enable observations of prestellar cores, protostellar envelopes, and protostellar shocks down to an angular resolution of a few hundreds au in the nearest star-forming regions. Subsequently, SKA-Mid Band 5 (covering the 4.6–15.4 GHz range) will allow for the exploration of this chemical complexity down to the 100 au scale within planet-forming regions.  
With these new facilities, multiple transitions of cyanopolyynes up to HC$_{11}$N, and their precursors C$_4$H and C$_6$H, will be detectable in several sources, resolving emission both spatially and spectrally in just a few hours of observation for extended sources such as prestellar cores and protostellar shocks, and in 100-1000 hours for compact planet-forming disks. 
These future SKA observations, coupled with an exhaustive revision of the chemical reaction networks, will provide crucial new insights into the formation and destruction routes of these species and their possible role in promoting prebiotic chemistry.


\section{Conclusions}\label{sec:conclusions}

In this paper, we report new observations of several carbon chains conducted with the 100-m GBT. These observations include Ku-band data (13.5-15.4 GHz) targeting both the L1544 prestellar core and the IRAS 16293 protostellar envelope, along with additional X-band data (8.0-11.6 GHz) towards L1544.
The main results of the paper are:


\begin{itemize}
    \vspace{-5pt}
    \item[-] In L1544, we detect emission lines from C$_2$S, C$_3$S, C$_3$N, c-C$_3$H, C$_4$H and C$_6$H (with E$_{\rm up}$$\leq$2~K), in addition to those of HC$_3$N, HC$_5$N, HC$_7$N and HC$_9$N, already reported in \citet{Bianchi2023}. All the lines have FWHM around 0.4 km s$^{-1}$ and show similar double-peaked profiles, suggesting a common underlying gas component.
    \vspace{1pt}
    \item[-] In IRAS 16293, we detect c-C$_3$H and, for the first time, HC$_7$N (emission lines with E$_{\rm up}$$\leq$5~K). We also report, for the first time, upper limits for C$_6$H and HC$_9$N. The detected lines have FWHM$\leq$1 km s$^{-1}$ and they arise from the cold envelope. For c-C$_3$H emission, we have contribution from both the cold envelope and the red-shifted outflow that we spectrally disentangle.
    \vspace{1pt}
    \item[-] We refine the spectroscopic frequencies of several c-C$_3$H and C$_6$H transitions, thanks to the high spectral resolution of the observations (1.4 kHz). The shifts between calculated and observed frequencies are consistent in both L1544 and IRAS 16293. 
    \vspace{1pt}
    \item[-] We performed LTE analysis to derive column densities of the detected species, assuming gas temperatures in the range 5-12 K for L1544, and 5-20 K for IRAS 16293, respectively. All column density estimates obtained in this work are consistent with previous measurements, where available. The only exception is C$_4$H, for which we used the revised dipole moment reported by \citet{oyama2020}, that yields column densities lower by a factor $\sim$6 compared to previous estimates. 
    In IRAS16293 we also calculated, for the first time, the c-C$_3$H column density and abundance ($\sim$$10^{-10}$) in the outflow wing.
    \vspace{1pt}
    \item[-] IRAS 16293 is less rich in carbon chains compared to L1544, with all detected species (except for c-C$_3$H) showing column densities lower by a factor $\sim$10-100.
    Despite this difference, the ratios between polyynyl radicals (C$_{\rm 2n}$H) and cyanopolyynes (HC$_{\rm 2n+1}$N) are similar in both sources (i.e. C$_4$H:C$_6$H = 1:0.053 and HC$_5$N:HC$_7$N:HC$_9$N = 1:0.16:0.25 in L1544, and 1:0.048 and 1:0.14:<0.28 in IRAS 16293, respectively).
    When compared to other prestellar and protostellar sources at different evolutionary stages, cyanopolyynes and polyynyl radicals show the same exponentially decreasing trend, suggesting that their chemistry is likely the same across different cold environments.
    \vspace{1pt}
    \item[-] We used a gas-phase astrochemical model to reproduce the abundances and investigate the formation of carbon chains. The model employing A$_{\rm v}$=10, standard $\zeta_{\rm CR}$, and typical C/O ratio reproduces the c-C$_3$H abundance in both sources, and the C$_6$H abundance in L1544. Conversely, while HC$_5$N is well reproduced in IRAS 16293, it is underestimated by more than an order of magnitude in L1544. The discrepancy is even more pronounced for HC$_7$N, whose abundance is systematically underpredicted by about two orders of magnitude in both sources.
    \vspace{1pt}
    \item[-] Models reproduce the C$_6$H/C$_4$H and HC$_5$N/HC$_3$N ratios in both L1544 and IRAS 16293 for A$_{\rm v}$=10 and t=2-7$\times$10$^5$ years. However, the HC$_7$N/HC$_5$N and HC$_9$N/HC$_7$N ratios are systematically underestimated, likely due to incomplete chemical networks for large cyanopolyynes. 
   \vspace{-2pt}  
\end{itemize}

This work paves the way for future astrochemical studies with next-generation radio interferometers such as SKAO. 
By enabling high-angular resolution observations of complex carbon-bearing molecules, these facilities will allow us to probe chemical complexity down to 100 au scales within star-forming regions. Combined with updated chemical reaction networks, such observations will provide critical insights into the formation carbon chains and cyanopolyynes and their prebiotic role.

\section*{Acknowledgements}

We thank the anonymous referee for the instructive comments which improved the paper. 
The authors warmly acknowledge Angelique Kahle and Friedrich Wyrowski for providing the CO~(3-2) APEX data.
This project has received funding from the EC H2020 research and innovation
programme for: (i) the project "Astro-Chemical Origins” (ACO, No 811312), and (ii) the European Research Council (ERC) project “The Dawn of Organic
Chemistry” (DOC, No 741002).  CC, LP, and GS acknowledge 
the PRIN-MUR 2020  BEYOND-2p (Astrochemistry beyond the second period elements, Prot. 2020AFB3FX), the project ASI-Astrobiologia 2023 MIGLIORA
(Modeling Chemical Complexity, F83C23000800005), the INAF-GO 2024 fundings ICES, and the INAF-GO 2023 fundings
PROTO-SKA (Exploiting ALMA data to study planet forming disks: preparing the advent of SKA, C13C23000770005). 
LP, CC, and GS also acknowledge financial support
under the National Recovery and Resilience Plan (NRRP), Mission 4, Component 2, Investment 1.1, Call for tender No. 104 published on 2.2.2022 by
the Italian Ministry of University and Research (MUR), funded by the European Union – NextGenerationEU-Project Title 2022JC2Y93 Chemical Origins:
linking the fossil composition of the Solar System with the chemistry of protoplanetary disks – CUP J53D23001600006 – Grant Assignment Decree No.
962 adopted on 30.06.2023 by the Italian Ministry of Ministry of University and Research (MUR). 
GS also acknowledges support from the INAF-Minigrant 2023 TRIESTE ("TRacing the chemIcal hEritage of our originS: from proTostars to planEts”).
EB acknowledges the support from the Italian Ministry for Universities and Research under the Italian Science Fund (FIS 2 Call - Ministerial Decree No. 1236 of 1 August 2023) and the Next Generation EU funds within the National Recovery and Resilience Plan (PNRR), Mission 4 - Education and Research, Component 2 - From Research to Business (M4C2), Investment Line 3.1 - Strengthening and creation of Research Infrastructures, Project IR0000034 – “STILES - Strengthening the Italian Leadership in ELT and SKA". 
The Green Bank Observatory is a facility of the National Science Foundation operated under cooperative agreement by Associated Universities, Inc.
LG acknowledges support from the COST Action CA22133 PLANETS.
\\

\section*{Data Availability}

The data underlying this article are available in the NRAO archive at \url{https://data.nrao.edu/portal/#/}, and can be accessed with the project code.



\bibliographystyle{mnras}
\bibliography{Mybib} 




\appendix

\section{Incomplete ion-neutral chemistry in cyanopolyyne networks}

Astrochemical reaction networks are typically based on the two most widely used databases, \citep[Kinetic Database for Astrochemistry:][]{wakelam2024} and UDfA \citep[UMIST Database for Astrochemistry:][]{Umist2022}. The GRETOPABE network adopted in this work is also KIDA-based, updated with recent results from our group and others on selected iCOMs \cite[e.g.][]{loison2014interstellar,neufeld2015sulphur,balucani2015formation,vazart2016state,skouteris2018genealogical,vazart2020gas,codella2020seeds,blazquez2020gas,giani2023revised}. Nevertheless, for many complex species with limited experimental and theoretical studies, such as cyanopolyynes, the networks remain incomplete, leading to large uncertainties in model predictions. For example, as one moves from HC$_3$N to HC$_{11}$N, the number of known production reactions decreases drastically (from 46 to 27). While the neutral–neutral chemistry of HC$_5$N has recently been revised \citep{Giani2025}, the ion–neutral pathways remain poorly constrained. 
Ionic reactions thought to be relevant for HC$_\mathrm{n}$N formation, particularly at times <10$^5$ years, include the electron recombination processes HC$_\mathrm{n}$NH$^+$ + e$^-$ and H$_3$C$_\mathrm{n}$N$^+$ + e$^-$ \citep{2004Geppert,Loison2017}, yet the networks of the corresponding protonated precursors are largely incomplete. For instance, current databases list 28 formation reactions for HC$_3$NH$^+$, five for HC$_5$NH$^+$, and only four for HC$_7$NH$^+$ and HC$_9$NH$^+$. 
In the case of H$_3$C$_\mathrm{n}$N$^+$ ions, four reactions are listed for each species, but none have been studied theoretically or experimentally, and identical rates and products are assumed in each case.
In our network, the formation reactions of HC$_7$NH$^+$ and HC$_9$NH$^+$ from HC$_7$N$^+$ and HC$_9$N$^+$ are missing, even though these are among the key reactions responsible for the formation of HC$_3$N and HC$_5$N. While the most recent updates of the KIDA and UDfA databases now include these reactions, their rate coefficients remain uninvestigated. We therefore suggest revising both the formation reactions of HC$_{\rm n}$NH$^+$, namely, HC$_{\rm n}$N$^+$ + H$_2$, and those leading to HC$_{\rm n}$N$^+$ itself, such as C$_{\rm n}$H$_3^+$ + N and C$_{\rm n}$N$^+$ + H$_2$. The omission of these reactions, as well as the use of uncertain rate coefficients, may explain why our models underestimate the abundances of the larger cyanopolyynes. In particular, we expect that including these missing pathways could reduce the discrepancy of about one to two orders of magnitude between the model predictions and the observed abundances of the longer cyanopolyynes.
However, arbitrarily adding reactions without properly revising their products and rates would introduce even larger uncertainties. We therefore strongly recommend a systematic revision of the ionic chemistry of cyanopolyynes to make their formation networks more complete, a task we plan to address in future work.



\bsp	
\label{lastpage}
\end{document}